\documentclass[pre,aps,twocolumn,superscriptaddress]{revtex4-2}

\usepackage{graphicx}
\usepackage{amssymb,amsfonts,amsmath}
\usepackage{color}
\usepackage[hidelinks]{hyperref}
\usepackage{bbm}

\usepackage{tikz}
\usetikzlibrary{shapes}
\usetikzlibrary{patterns}
\usetikzlibrary{angles, quotes}
\definecolor{bostonuniversityred}{rgb}{0.8, 0.0, 0.0}
\definecolor{chromeyellow}{rgb}{1.0, 0.65, 0.0}

\newcommand{\rv}{{\mathbf r}}
\renewcommand{\vec}{\mathbf}
\newcommand{\rhotwo}{\rho^{(2)}}

\usepackage{amssymb}
\usepackage{bbold}
\newcommand{\unit}{\vec{e}}

\begin{document}


\title{First-principles superadiabatic theory for the dynamics of inhomogeneous fluids}


\author{S.~M. Tschopp}
\author{J.~M. Brader}
\affiliation{Department of Physics, University of Fribourg, CH-1700 Fribourg, Switzerland}

\date{\today}

\begin{abstract}
For classical many-body systems subject to Brownian 
dynamics we develop a superadiabatic dynamical density functional theory (DDFT) for the description of inhomogeneous fluids out-of-equilibrium.
By explicitly incorporating the dynamics of the inhomogeneous two-body correlation functions we obtain superadiabatic forces directly from the microscopic interparticle interactions.  
We demonstrate the importance of these nonequilibrium forces for an accurate description of the one-body density by numerical implementation of our theory for three-dimensional hard-spheres in a time-dependent planar potential. 
The relaxation of the one-body density in  superadiabatic-DDFT is found to be slower 
than that predicted by standard adiabatic DDFT and significantly improves the agreement with Brownian dynamics simulation data. 
We attribute this improved performance to the correct treatment of structural relaxation within 
the superadiabatic-DDFT. 
Our approach provides fundamental insight into the underlying structure of dynamical density functional theories and makes possible the study of situations for which standard approaches fail. 

\end{abstract}

\maketitle


\section{Introduction}

The density functional theory (DFT) provides an exact framework for the study of classical many-body systems in equilibrium under the influence of external fields \cite{Evans79,Evans92}. 
Minimization of the grand potential functional with respect 
to the one-body density yields both thermodynamic quantities and the spatial correlation functions which characterize the average particle microstructure. 
Although the exact form of the grand potential functional is unknown in general, the physically intuitive nature of the formalism facilitates the construction of useful approximation schemes for a wide variety of interesting model systems \cite{Evans92,RothReview,WittmannFMT,tschopp1}. 
As a consequence, DFT has been employed to study many fundamental phenomena \cite{Evans92} from freezing \cite{Lowen,oxtoby} 
and liquid-crystal phases \cite{WittmannFMT} to wetting and drying at interfaces \cite{EvansWilding}.

Given that DFT provides a solid framework for the study of inhomogeneous fluids in
equilibrium, it is natural to seek a nonequilibrium generalization to treat situations involving a time-dependent external field. 
The first step is to specify the level of resolution desired for the description of the microscopic dynamics. 
The treatment of many-body Newtonian systems is theoretically challenging; the inclusion of inertia leads almost inevitably to hydrodynamic-type approaches lacking spatial resolution on the particle scale.  
This poses difficulties for the detailed investigation of dynamical phenomena within the interface or in confined fluids. 
Fortunately, for many situations Brownian dynamics is the appropriate choice on the time- and length-scales of interest, particularly within the field of soft matter physics. 
In this case, a simple theory for the time-dependent one-body density can be constructed by assuming that the average particle 
current arises from the gradient of a local chemical potential
\cite{Evans79,Dieterich1,Dieterich2,Marconi98}. 
Relating this to the functional derivative of the
equilibrium free energy and using the continuity equation then leads to 
the standard dynamical density functional theory (DDFT) for 
Brownian dynamics. 
This theory has been applied with considerable success to 
many nonequilibrium situations, such as diffusion in periodically varying external fields 
\cite{Marconi98,rex}, sedimentation in a gravitational field 
\cite{sedimentation1,sedimentation2} and phase separation dynamics \cite{ArcherEvansDDFT,phase_cavity_Archer}. 
(For a more complete overview of the field we refer the reader reference \cite{biglongreview}.)

During the last few decades the DDFT has been rederived 
using various methods: 
directly from a stochastic Langevin equation 
\cite{Marconi98}, by coarse-graining the many-body 
Smoluchowski equation \cite{ArcherEvansDDFT},  
or by using the projection operator formalism 
\cite{projection}. 
However, despite the insight provided by alternative derivations of the DDFT, the
final equation of motion for the density is always the same. 
No practical guidance is provided for making a systematic,  or indeed any, improvement to the underlying structure of the theory. 
The necessity of such a step is apparent when considering the
asymmetry between the treatment of spatial and temporal degrees of freedom within DDFT. 
The intricate nonlocal spatial structure of the equilibrium free energy functional stands in stark contrast to the simple, time-local treatment of the dynamics.  
This suggests that there is much room for improvement, particularly in view of 
memory-function approaches, such as mode-coupling theory 
\cite{modecoupling}. 
This issue is addressed by the power functional theory (PFT), a nonequilibrium generalization of DFT based on the minimization of a free power functional 
\cite{PFToriginal,SchmidtYellow}. 
In addition to motivating phenomenological approximations 
beyond standard DDFT \cite{velocitygradientPFT,structuraldrivenPFT,SchmidtYellow}, the formal structure of PFT has served to focus simulation based studies and facilitated interpretation of the data 
\cite{Fortini,flowandstructurePFT,structuraldrivenPFT,customflowPFT}. 
However, what remains to be found is a first-principles approximation which makes a direct connection between the free 
power and the microscopic interparticle interactions.

At the core of DDFT lies the so-called adiabatic approximation \cite{Marconi98,SchmidtYellow}. 
In the present context this refers to the assumption that 
the true nonequilibrium pair correlations 
are equal to those of a fictitious equilibrium with a density distribution given by the instantaneous density of the nonequilibrium system. 
This tethers the time-evolution of the two-body density, $\rhotwo(\rv_1,\rv_2,t)$, to that 
of the one-body density $\rho(\rv_1,t)$ and ignores the fact that the former has 
its own dynamics and relaxation time.  
While this assumption provides reasonable qualitative results for relaxational dynamics close to 
equilibrium, it breaks down when internal relaxation times become long, as is the case for 
high density liquids \cite{modecoupling}, systems with strong interparticle attraction (e.g.~gels) \cite{CatesSoftBook}, 
or when the theory is extended to treat driven systems with nonconservative forces \cite{BraderKruger,AerovKrueger,BraderScacchi}. 
In bulk systems under external driving (e.g.~shear) it is well-known that the distorted 
two-body correlations encode essential information about the 
average forces acting within the fluid \cite{BraderReview}. 
It is thus to be expected that the inhomogeneous two-body correlations 
will play a similarly important role for spatially inhomogeneous systems out-of-equilibrium.

The lesson to be taken from the examples given above is that any physical mechanism which can only be resolved on the two-body level (or above) cannot be accounted for by standard DDFT. 
What is required is a more general theoretical approach to interacting Brownian systems which provides the next-leading order beyond the adiabatic
approximation. 
This improved theory would not only give fundamental insight into the basic stucture of DDFT, but would also 
make possible applications where the standard DDFT fails. 
The clearest way forwards 
is to consider explicitly the dynamics of the two-body 
inhomogeneous correlation functions, releasing them from the adiabatic constraint that 
they are determined entirely by the instantaneous one-body density. 
Treating the coupled dynamics of the one- and two-body densities would automatically generate an equation of motion for the one-body density exhibiting, albeit implicitly, the expected temporal nonlocality; a true superadiabatic-DDFT.
Within such an approach the two-body correlations would play the role of auxilliarly functionals which introduce memory 
into the dynamics of the one-body density. 

In this paper we both derive and implement the desired 
superadiabatic-DDFT. This incorporates memory effects and 
is capable of predicting superadiabatic forces 
directly from the underlying interparticle interactions. 
By explicitly addressing the dynamics of the two-body correlation functions our theory captures physical 
mechanisms absent from the standard DDFT. 
We illustrate this by implementing the theory for a 
system of hard-spheres in two different planar external fields.

\section{Theory}\label{sec:Theory}

\subsection{One-body density in- and out-of-equilibrium}

For a system of $N$ interacting Brownian particles 
the time-evolution of the configurational probability density, $P(\rv^N\!,t)$, where $\rv^N$ represents the set of all coordinates, $\{\rv_1 \ldots \rv_N\}$, is given by the Smoluchowski 
equation   
\begin{multline}\label{smoluchowski}
\frac{1}{D_0}\frac{\partial P(\rv^N\!,t)}{\partial t}
=
\sum_{i=1}^{N}
\nabla_{\vec{r}_i} \cdot \Big( P(\rv^N\!,t)\big(
\nabla_{\vec{r}_i} \ln (P(\rv^N\!,t)) \Big. \\ \Big. 
+ \nabla_{\vec{r}_i} \beta U(\rv^N\!,t)
\big) \Big),
\end{multline}
where $\beta\!=\!(k_BT)^{-1}$ and $D_0$ is the bare diffusion coefficient.  
For systems with pairwise interactions 
the total potential energy is given by 
\begin{equation}
U(\rv^N\!,t) = \sum_{i<j}^{N}\phi(r_{ij}) 
+ \sum_{i=1}^{N}V_{\text{ext}}(\rv_i,t),
\end{equation}
where $\phi$ is the pair potential, $r_{ij}=|\vec{r}_i-
\vec{r}_j|$ and 
$V_{\text{ext}}$ is a time-dependent external 
potential. 
For the present work we do not consider the influence 
of nonconservative forces (e.g.~shear), but these can be easily incorporated  into equation \eqref{smoluchowski} if required (see section 
\ref{discussion} for a more detailed discussion of this point).

Integrating equation \eqref{smoluchowski} over $N\!-\!1$ particle coordinates generates a formally exact equation of motion for the one-body density
\begin{align} \label{one-body exact}
\frac{1}{D_0}  \frac{\partial \rho(\vec{r}_1,t)}{\partial t} =& \,\nabla_{\vec{r}_1} \!\!\cdot \!\Bigg(\!\nabla_{\vec{r}_1} \rho(\vec{r}_1,t) + \rho(\vec{r}_1,t) \nabla_{\vec{r}_1} \beta V_{\text{ext}}(\vec{r}_1,t) \notag\\
&+ \int d \vec{r}_2 \, \rho^{(2)}(\vec{r}_1,\vec{r}_2,t) \nabla_{\vec{r}_1} \beta \phi(r_{12}) \!\Bigg)
\\
\equiv& -\nabla_{\vec{r}_1} \cdot \vec{j}(\vec{r}_1,t),
\label{one-body current}
\end{align}
where $\vec{j}$ is the one-body current. 
Confining fields or substrates are 
often modelled using an external potential which jumps discontinuously 
between zero and infinity as a function of position, 
e.g.~for a planar hard-wall or a spherical confining cavity.    
In such cases it is preferable to remove the term 
$\rho(\vec{r}_1,t) \nabla_{\vec{r}_1} \beta V_{\text{ext}}(\vec{r}_1,t)$ from equation \eqref{one-body exact} and 
rather account for the external potential using a zero-flux condition on the boundary, 
namely $\vec{j}\cdot \hat{\textbf{n}}\!=\!0$, where 
$\hat{\textbf{n}}$ is a unit vector normal the substrate 
or confining surface \cite{Cichocki}.   

In equilibrium the 
one-body current vanishes and we obtain the first-order Yvon-Born-Green (YBG) equation 
\begin{multline} \label{YBG 1}
\nabla_{\vec{r}_1} \rho_{\text{eq}}(\vec{r}_1) + \rho_{\text{eq}}(\vec{r}_1) \nabla_{\vec{r}_1} \beta V_{\text{ext}}(\vec{r}_1)\\ 
+ \int d \vec{r}_2 \, \rho_{\text{eq}}^{(2)}(\vec{r}_1,\vec{r}_2; [\rho_{\text{eq}}]) \nabla_{\vec{r}_1} \beta \phi(r_{12}) = 0, 
\end{multline}
which expresses the balance between 
Brownian, external and internal forces \cite{Hansen06}. 
The notation employed in equation \eqref{YBG 1} makes explicit the fact that the equilibrium two-body density 
is a functional of the one-body density \cite{tschopp1,tschopp2}. 
In this sense, the YBG equation presents a formally closed theory for $\rho_{\text{eq}}$. The difficulty, of course, lies 
in actually obtaining this two-body density functional.  
Of the various methods available the most effective are based on the inhomogeneous Ornstein-Zernike (OZ) equation
\begin{multline} \label{oz}
h_{\text{eq}}(\rv_1,\rv_2; [\rho_{\text{eq}}]) = c^{(2)}_{\text{eq}}(\rv_1,\rv_2; [\rho_{\text{eq}}]) \\ + \int\! d\rv_3\, h_{\text{eq}}(\rv_1,\rv_3; [\rho_{\text{eq}}])\, \rho_{\text{eq}}(\rv_3) \, c^{(2)}_{\text{eq}}(\rv_3,\rv_2; [\rho_{\text{eq}}]),
\end{multline}
where $c^{(2)}_{\text{eq}}$ is the two-body direct correlation function 
and $h_{\text{eq}}$ is the total correlation function, related to the 
two-body density according to 
\begin{equation} \label{rho2 definition}
h_{\text{eq}}(\rv_1, \rv_2; [\rho_{\text{eq}}]) = \frac{\rho^{(2)}_{\text{eq}}(\rv_1, \rv_2; [\rho_{\text{eq}}])}{\rho_{\text{eq}}(\rv_1) \rho_{\text{eq}}(\rv_2)} -1.
\end{equation} 
Since both $c^{(2)}_{\text{eq}}$ and $h_{\text{eq}}$ are unknown in equation \eqref{oz} a second, independent, 
relation is required. 
One possibility is to employ a closure in the spirit of liquid-state integral equation theory, such as the Percus-Yevick or Hyper-Netted-Chain approximations \cite{AttardBook}. Another option is to calculate $c^{(2)}_{\text{eq}}$ 
from a second functional derivative of the excess (over ideal) Helmholtz free energy functional, 
\begin{equation} \label{c2 functional}
c^{(2)}_{\text{eq}}(\rv_1,\rv_2; [\rho_{\text{eq}}])=-{\frac{\delta^2 \beta F^{\,\text{exc}}[\rho]}{\delta\rho(\rv_1) \delta\rho(\rv_2)}}\bigg\rvert_{\rho_{\text{eq}}}, 
\end{equation}
and then substitute this into equation \eqref{oz}. 
In any case, the equilibrium YBG theory described above holds regardless of the specific scheme chosen to obtain the equilibrium two-body density as a functional of the inhomogeneous one-body density.

Taking the closed set of equations \eqref{YBG 1},\eqref{oz},\eqref{rho2 definition} and \eqref{c2 functional} yields the so-called force-DFT, recently presented and discussed in detail in Reference \cite{tschopp3}.
This provides a method to obtain the density profile 
from a given excess Helmholtz free energy functional. 
If the latter is known 
exactly, then the force-DFT is equivalent to the standard 
DFT approach based on minimization of the grand potential functional, $\Omega[\rho]$, see Appendix 
\ref{appendix potential-DFT}. 
Employing an approximate excess free energy will lead 
to differences between the density profiles predicted by standard DFT and force-DFT \cite{tschopp3}. 
The main technical challenge when implementing force-DFT is solution of the inhomogeneous OZ equation \eqref{oz}, due to the appearance of the spatially varying one-body density within the integral term. 
In contrast to bulk systems, for which Fourier transformation 
reduces the homogeneous OZ equation to an algebraic relation,
numerical solution of \eqref{oz} is generally prohibitive. 
Nevertheless, as shown in \cite{AttardBook,AttardSpherical,tschopp1,tschopp2}, it can be achieved in situations of high symmetry, namely in planar and spherical geometry, using the appropriate integral transform (Hankel and Legendre, respectively).

\subsection{Two-body density in- and out-of-equilibrium}

Integrating the Smoluchowski equation \eqref{smoluchowski} over $N\!-\!2$ particle coordinates yields the following exact equation of motion for the two-body density
\begin{align} \label{two-body exact}
\frac{1}{D_0} \, \frac{\partial \rho^{(2)}(\vec{r}_1,\vec{r}_2,t)}{\partial t} &\!=\! \sum_{i=1,2} \nabla_{\vec{r}_i} \cdot \Bigg(\! \nabla_{\vec{r}_i} \rho^{(2)}(\vec{r}_1,\vec{r}_2,t) 
\notag\\
&\hspace*{-1cm}+ \rho^{(2)}(\vec{r}_1,\vec{r}_2,t) \nabla_{\vec{r}_i} \beta \big(V_{\text{ext}}(\vec{r}_i)+\phi(r_{12})\big) 
\notag\\
&\hspace*{-1cm}+ \!\int d \vec{r}_3 \, \rho^{(3)}(\vec{r}_1,\vec{r}_2,\vec{r}_3,t) \nabla_{\vec{r}_i} \beta \phi(r_{i3}) \!\Bigg)
\\
&\hspace*{-1cm}\equiv-\sum_{i=1,2} \nabla_{\vec{r}_i}\cdot\textbf{j}_i^{(2)}(\textbf{r}_1,\textbf{r}_2,t),
\label{two-body current}
\end{align}
where $\rho^{(3)}$ is the nonequilibrium three-body 
density and $\textbf{j}_i^{(2)}$ is the two-body current at coordinate $\textbf{r}_i$. 
Equation \eqref{two-body current} shows that 
the time-evolution of the two-body density requires information about the divergence of the two-body current at both of its spatial arguments, $\textbf{r}_1$ and $\textbf{r}_2$. 
An important feature of equation \eqref{two-body exact} is 
that the pair potential appears not only in integrated 
form, describing mediated forces, but also in `bare' form, 
as a direct generator of force between the density at 
coordinate $\vec{r}_1$ and that at coordinate $\vec{r}_2$. 
In systems for which the pair potential diverges at small 
separations, the direct interaction force ensures that 
$\rho^{(2)}(\vec{r}_1,\vec{r}_2,t)\rightarrow 0$ as 
$r_{12}\rightarrow 0$, reflecting the impossibility of full particle overlap.  
In the case of hard-spheres the pair potential generates 
an infinitely repulsive force on the excluded volume 
`contact shell'. It is then preferable to omit the singular term 
$\rho^{(2)}(\vec{r}_1,\vec{r}_2,t) \nabla_{\vec{r}_i} \beta\phi(r_{12})$ from equation 
\eqref{two-body exact} and rather account for this by imposing the zero-flux condition 
$\textbf{j}^{(2)}\!\cdot\hat{\textbf{n}}_{\,r_{12}}\!=\!0$, where $\hat{\textbf{n}}_{\,r_{12}}\!\equiv\!(\textbf{r}_1-\textbf{r}_2)/r_{12}$ is a unit vector \cite{Cichocki}. 
This guarantees that the exact core-condition, $\rho^{(2)}(\vec{r}_1,\vec{r}_2,t)\!=\! 0$ for $r_{12}\!<\!d$, is satisfied for all times, where $d$ is the hard-sphere diameter.

In equilibrium the two-body current vanishes independently for each value of the coordinate index $i\!=\!1,2$. 
This yields the second-order Yvon-Born-Green (YBG2) equation 
\cite{Hansen06}
\begin{align} \label{YBG 2}
\nabla_{\vec{r}_i} &\rho_{\text{eq}}^{(2)}(\vec{r}_1,\vec{r}_2; [\rho_{\text{eq}}]) \\
&+ \rho_{\text{eq}}^{(2)}(\vec{r}_1,\vec{r}_2; [\rho_{\text{eq}}]) \nabla_{\vec{r}_i} 
\beta \big(V_{\text{ext}}(\vec{r}_i)+\phi(r_{12})\big)
\notag\\
&+ \int d \vec{r}_3 \, \rho_{\text{eq}}^{(3)}(\vec{r}_1,\vec{r}_2,\vec{r}_3; [\rho_{\text{eq}}]) \nabla_{\vec{r}_i} \beta \phi(r_{i3}) \,\, =0, \notag
\end{align}
where $\rho_{\text{eq}}^{(3)}$ 
is the inhomogeneous equilibrium three-body density.
If this is a known functional of the one-body density, then equation \eqref{YBG 2} provides, in principle, a means to 
obtain the inhomogeneous two-body density. 
However, in practice, attempts to implement such a scheme are generally limited to bulk systems at low density, 
where the Kirkwood superposition approximation can be applied \cite{Hansen06,mcquarrie}.

\subsection{Superadiabatic-DDFT}\label{subsec:superadiabatic}

The two exact equations \eqref{one-body exact} and \eqref{two-body exact} form a system of 
first-order differential equations for the coupled dynamics 
of the one- and two-body densities. 
However, to close the theory, equation \eqref{two-body exact} requires an approximation to the nonequilibrium three-body density. 
In this subsection we will employ the YBG2 equation \eqref{YBG 2} to approximate the three-body integral in terms of known equilibrium two-body functionals. 
This will enable us to explicitly treat the dynamics of the 
nonequilibrium two-body density and generate, via equation \eqref{one-body exact}, a closed dynamical theory 
for the one-body density which goes beyond standard DDFT. 

A key concept in addressing the dynamics of inhomogeneous fluids is the `adiabatic system'. This is a fictitious equilibrium with a one-body density equal to the instantaneous density of the real nonequilibrium system of interest. 
The adiabatic system has the same interaction potential 
as the real system, but is subject to a modified external field, chosen to provide a match between the adiabatic 
and real one-body densities. 
This naturally implies that the adiabatic external field is time-dependent and that it must be recalculated at each 
time-step.  
We have already established that the equilibrium two-body density is a functional of $\rho_{\text{eq}}$. 
If we evaluate this functional using a nonequilibrium 
one-body density, then we obtain the two-body density of the adiabatic system
\begin{equation} \label{rho2 ad}
\rho^{(2)}_{\text{ad}}(\textbf{r}_1,\textbf{r}_2,t)
\equiv
\rho^{(2)}_{\text{eq}}(\textbf{r}_1,\textbf{r}_2;
[\rho(\textbf{r},t)]).
\end{equation}
Similarly, we can define two additional quantities 
which will be of use later on. 
The adiabatic three-body density is defined according to 
\begin{equation} \label{rho3 ad}
\rho^{(3)}_{\text{ad}}(\textbf{r}_1,\textbf{r}_2,\textbf{r}_3,t)
\equiv
\rho^{(3)}_{\text{eq}}(\textbf{r}_1,\textbf{r}_2,\textbf{r}_3;
[\rho(\textbf{r},t)]).
\end{equation}
Evaluating the equilibrium YBG equation 
\eqref{YBG 1} at the nonequilibrium instantaneous density 
generates the fictitious external force field required to stabilize the adiabatic system. 
This external force is given by
\begin{multline} \label{V ad} 
-\nabla_{\vec{r}_1} V_{\text{ad}}(\vec{r}_1,t) 
\equiv
k_BT\,\nabla_{\vec{r}_1} \ln \rho(\vec{r}_1,t) 
\\ 
+ \int d \vec{r}_2 \, 
\frac{\rho_{\text{ad}}^{(2)}(\vec{r}_1,\vec{r}_2,t)}
{\rho(\vec{r}_1,t)}
\nabla_{\vec{r}_1}\phi(r_{12}), 
\end{multline}
where $V_{\text{ad}}$ is an adiabatic potential, defined by 
equation \eqref{V ad} up to a physically irrelevant additive 
constant.  

Substitution of the adiabatic two-body density into 
the exact equation 
\eqref{one-body exact} yields a closed equation of motion for the nonequilibrium one-body density
\begin{align} \label{force-DDFT}
\!\!\!\frac{1}{D_0} \frac{\partial \rho(\vec{r}_1,t)}{\partial t} =& \nabla_{\vec{r}_1} \!\!\cdot\! \Bigg(\! \nabla_{\vec{r}_1} \rho(\vec{r}_1,t) + \rho(\vec{r}_1,t) \nabla_{\vec{r}_1} \beta V_{\text{ext}}(\vec{r}_1,t) \notag\\
&+ \int d \vec{r}_2 \, \rho^{(2)}_{\text{ad}}(\vec{r}_1,\vec{r}_2,t) \nabla_{\vec{r}_1} \beta \phi(r_{12}) \!\Bigg). 
\end{align}
This adiabatic dynamical equation (`force-DDFT') was recently 
presented and discussed in detail in Reference \cite{tschopp3}. The required adiabatic two-body density functional is obtained 
by substituting $\rho(\textbf{r}_1,t)$ into the 
equilibrium two-body direct correlation function 
\eqref{c2 functional}, solving the inhomogeneous OZ equation  \eqref{oz} and then employing the definition 
\eqref{rho2 definition}. 
The interaction potential appears explicitly in equation 
\eqref{force-DDFT} via the integral term, but also 
implicitly through the excess Helmholtz free energy functional used to evaluate the two-body 
direct correlation function \eqref{c2 functional}. 
In equilibrium the one-body current vanishes and equation 
\eqref{force-DDFT} reduces to the force-DFT outlined earlier. 

The force-DDFT \eqref{force-DDFT} is an approximate equation of motion for the nonequilibrium one-body density, obtained by making an adiabatic approximation to the two-body density. To be clear about our terminology, and to remain as consistent as possible with existing literature, we will henceforth refer to the force-DDFT \eqref{force-DDFT} as generating `adiabatic dynamics on the one-body level'. 
The force-DDFT is different in spirit from the standard formulation of DDFT (recalled in Appendix \ref{appendix standard DDFT}) in 
which interaction forces are generated using the one-body direct correlation function \cite{Marconi98,ArcherEvansDDFT}.
These two approaches become equivalant only in the case that the excess Helmholtz free energy is known exactly. 
For further details regarding the force-DDFT we refer the reader to Reference \cite{tschopp3}.

To go beyond adiabatic dynamics on the one-body level 
a closed equation of motion for the 
two-body density is required. 
We thus make the following approximation
\begin{align} \label{triple integral ad}
\int d \vec{r}_3 \, &\rho^{(3)}(\vec{r}_1,\vec{r}_2,\vec{r}_3,t) \nabla_{\vec{r}_i} \phi(r_{i3}) \notag \\
& \!\!\!\!\!\!\approx \int d \vec{r}_3 \,  \rho^{(3)}_{\text{ad}}(\vec{r}_1,\vec{r}_2,\vec{r}_3,t)  \nabla_{\vec{r}_i} \phi(r_{i3}),
\end{align}
where the adiabatic three-body density is given by equation \eqref{rho3 ad}.
We then use the equilibrium YBG2 equation \eqref{YBG 2} and express 
the three-body integral \eqref{triple integral ad} in terms of known adiabatic two-body functionals,  
\begin{align} \label{apply ybg2}
&\int d \vec{r}_3 \,  \rho^{(3)}_{\text{ad}}(\vec{r}_1,\vec{r}_2,\vec{r}_3,t)  \nabla_{\vec{r}_i} \phi(r_{i3})
\\ 
&\stackrel{\text{YBG2}}{=}\! 
-k_BT\,\nabla_{\vec{r}_i} \rho^{(2)}_{\text{ad}}(\vec{r}_1,\vec{r}_2,t) - \rho^{(2)}_{\text{ad}}(\vec{r}_1,\vec{r}_2,t) \nabla_{\vec{r}_i} V_{\text{ad}}(\vec{r}_i,t) \notag \\
&\quad\quad \!- \rho^{(2)}_{\text{ad}}(\vec{r}_1,\vec{r}_2,t) \nabla_{\vec{r}_i} \phi(r_{12}),
\notag
\end{align}
where the adiabatic two-body density is given by 
\eqref{rho2 ad} and the gradient of the adiabatic potential is obtained from equation \eqref{V ad}.
Substitution of equation \eqref{apply ybg2} into equation \eqref{two-body exact} yields
\begin{align} \label{two body adiabatic}
&\frac{1}{D_0} \, \frac{\partial \rho^{(2)}(\vec{r}_1,\vec{r}_2,t)}{\partial t} = \\
&\sum_{i=1,2} \nabla_{\vec{r}_i} 
\!\cdot\! \Bigg( \!\nabla_{\vec{r}_i} \rho^{(2)}_{\text{\,sup}}(\vec{r}_1,\vec{r}_2,t)
\!+\! \rho^{(2)}_{\text{\,sup}}(\vec{r}_1,\vec{r}_2,t) \nabla_{\vec{r}_i} \beta \phi(r_{12}) \notag\\
&\!\!\!+\! \rho^{(2)}(\vec{r}_1,\vec{r}_2,t) \nabla_{\vec{r}_i} \beta V_{\text{ext}}(\vec{r}_i)
\!-\! \rho^{(2)}_{\text{ad}}(\vec{r}_1,\vec{r}_2,t) \nabla_{\vec{r}_i} \beta V_{\text{ad}}(\vec{r}_i,t) \!\!\Bigg) , \notag
\end{align}
where we have defined the superadiabatic contribution to the two-body density according to 
\begin{equation} \label{rho sup}
\rho^{(2)}_{\, \text{sup}}(\vec{r}_1,\vec{r}_2,t) \!\equiv\! \rho^{(2)}(\vec{r}_1,\vec{r}_2,t)-\rho^{(2)}_{\text{ad}}(\vec{r}_1,\vec{r}_2,t).
\end{equation}
The pair of coupled equations \eqref{one-body exact} and \eqref{two body adiabatic} form a closed set, which we will henceforth refer to as `superadiabatic-DDFT'. 
As with the force-DDFT discussed earlier, the adiabatic 
two-body density functional, $\rho^{(2)}_{\text{ad}}$, is calculated at each time-step 
by using the instantaneous density to evaluate the 
equilibrium two-body direct correlation function 
\eqref{c2 functional}, solving the inhomogeneous OZ equation  \eqref{oz} and then employing the definition 
\eqref{rho2 definition}. The gradient of the adiabatic potential 
is evaluated using equation \eqref{V ad}.
In equilibrium both the the one- and two-body currents vanish. 
From the second equation of superadiabatic-DDFT, 
equation \eqref{two body adiabatic}, it can be deduced that the two-body density reduces to the equilibrium form determined by equations \eqref{oz}-\eqref{c2 functional}.  
Consequently, the first equation of 
superadiabatic-DDFT, equation \eqref{one-body exact}, predicts 
the equilibrium one-body density profile of force-DFT. 

Equations \eqref{one-body exact} and \eqref{two body adiabatic} determine the simultaneous time-evolution of the one- and two-body densities.
These equations are both local in time. 
However, there is a fundamental difference between a single 
time-local differential equation, 
such as the standard or force-DDFT, 
and the new superadiabatic-DDFT.   
If we regard the nonequilibrium two-body density as an auxilliarly function, 
then its formal elimination from the system of equations would result in a stand-alone equation of motion for the one-body density involving a time-integral over the history 
of $\rho(\vec{r},t)$. 
Although an explicit realization of this elimination is 
prohibited by the complexity of the adiabatic functionals, our approach nevertheless contains an implicit mechanism 
for the generation of memory effects in the dynamics of the one-body density, without the need to explicitly approximate a memory kernel. 
The label `superadiabatic' used to describe our theory thus refers to the equation of motion for the 
one-body dynamics; employing adiabatic dynamics on the two-body level generates superadiabatic dynamics on the 
one-body level.   
The superadiabatic character of our approach can be made 
clearer if we rewrite the exact one-body equation 
of motion \eqref{one-body exact} to show more explicitly 
the various forces acting in the system, namely  
\begin{align} \label{one-body exact forces}
\frac{k_BT}{D_0} \, \frac{\partial \rho(\vec{r}_1,t)}{\partial t} &= -\nabla_{\vec{r}_1} \!\cdot\! \Big(
\vec{f}_{\,\text{Br}}(\vec{r}_1,t) \\
&+ 
\vec{f}_{\,\text{ext}}(\vec{r}_1,t) 
+ 
\vec{f}_{\,\text{ad}}(\vec{r}_1,t)
+ 
\vec{f}_{\,\text{sup}}(\vec{r}_1,t)
\Big).
\notag
\end{align}
The force densities generated by Brownian motion and the external field are given by
\begin{align}
\vec{f}_{\,\text{Br}}(\vec{r}_1,t) &= -k_B T\,\nabla_{\vec{r}_1} \rho(\vec{r}_1,t),
\notag\\
\vec{f}_{\,\text{ext}}(\vec{r}_1,t) &= -\rho(\vec{r}_1,t)\nabla_{\vec{r}_1} V_{\text{ext}}(\vec{r}_1,t), 
\notag
\end{align}
while the interparticle interactions appear via the adiabatic 
and superadiabatic forces, given by
\begin{align}
\vec{f}_{\,\text{ad}}(\vec{r}_1,t) &= -\!\!\int \! d \vec{r}_2 \, \rho^{(2)}_{\text{ad}}(\vec{r}_1,\vec{r}_2,t) \nabla_{\vec{r}_1} \phi(r_{12}) ,\label{f_ad}\\
\vec{f}_{\,\text{sup}}(\vec{r}_1,t) &= -\!\!\int \! d \vec{r}_2 \, \rho^{(2)}_{\, \text{sup}}(\vec{r}_1,\vec{r}_2,t) \nabla_{\vec{r}_1} \phi(r_{12}).
\label{f_sup}
\end{align}
The force-DDFT, equation \eqref{force-DDFT}, is recovered by setting the superadiabatic force equal to zero. 
From equations \eqref{rho sup} and 
\eqref{f_sup} it is clear that the superadiabatic force arises from differences between the nonequilibrium and the adiabatic two-body densities \cite{Fortini}. 
Equation \eqref{two body adiabatic} enables these differences to be calculated directly from the microscopic interparticle 
interactions.

\section{Application to three-dimensional hard-spheres in planar geometry}\label{sec:application hs}

\subsection{Analytics}
\begin{figure}
\begin{minipage}[t]{0.4\textwidth}
\hspace*{-0.5cm}
\begin{tikzpicture}
\coordinate (cross) at (2,0);
\coordinate (wall upper pt) at (-3,4);
\coordinate (wall lower pt) at (-3,-1);
\coordinate (wall size) at (-1,5);
\node[draw, circle] (particle 1) at (0,0) {1};
\node[draw, circle] (particle 2) at (2,3) {2};
\node (r spherical) at (1,1.5) {$r_{12}$};
\node[bostonuniversityred] (r cylindrical) at (2,1.5) {$r_2$};
\node[bostonuniversityred] (z1 label) at (-1.5,0) {$z_1$};
\node[bostonuniversityred] (z2 label) at (-0.5,3) {$z_2$};
\draw[-, >=latex, bostonuniversityred, line width=0.9,] (-3,0) -- (z1 label);
\draw[-, >=latex, bostonuniversityred, line width=0.9] (z1 label) -- (particle 1);
\draw[-, >=latex, bostonuniversityred, line width=0.9,] (-3,3) -- (z2 label);
\draw[-, >=latex, bostonuniversityred, line width=0.9,] (z2 label) -- (particle 2);
\draw[dashed, >=latex] (particle 1) -- (r spherical);
\draw[dashed, >=latex] (r spherical) -- (particle 2);
\draw[-, >=latex, bostonuniversityred, line width=0.9,] (particle 2) -- (r cylindrical);
\draw[-, >=latex, bostonuniversityred, line width=0.9,] (r cylindrical) -- (cross);
\draw[dashed, >=latex] (particle 1) -- (cross);
\draw[-, >=latex] (wall lower pt) -- (wall upper pt);
\fill[pattern=north west lines] (wall lower pt) rectangle ++(wall size);
\end{tikzpicture}
\end{minipage}
\hspace*{2cm}
\caption{\textbf{Sketch of the planar geometry.}  
Two-body correlation functions are conveniently expressed 
using cylindrical coordinates. 
We choose $z_1$ as the axis of our cylindrical coordinate system ($r_1$ is thus implicitly fixed equal to zero).
}
\label{sketch_planar}
\end{figure}
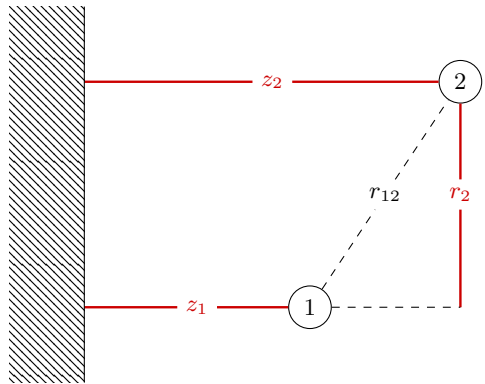

To investigate the predictions of superadiabatic-DDFT we will focus on three-dimensional systems 
in planar geometry, where the external field (and thus the one-body density) vary as a function of a single cartesian coordinate, taken here as the z-axis. 
In this particular geometry the inhomogeneous two-body correlation functions have cylindrical symmetry and require as input three scalar variables, $z_1$, $z_2$ and the cylindrical radial 
distance, $r_2$ (as illustrated in Fig.~\ref{sketch_planar}). 
We choose $z_1$ to be coincident with the 
$z$-axis of the cylindrical coordinate system 
\cite{AttardBook}. 
From here on we will focus exclusively on the 
three-dimensional hard-sphere system and 
set $D_0$, $k_B T$ and the hard-sphere diameter, $d$, equal to unity, 

In the chosen geometry the one-body equation of motion \eqref{one-body exact forces} takes the following form
\begin{align} \label{one body planar}
\frac{\partial \rho(z_1,t)}{\partial t} = -\frac{\partial}{\partial z_1} \Big( 
&f_{\text{Br}}(z_1,t) 
+ f_{\text{ext}}(z_1,t) 
\\
&+ f_{\text{ad}}(z_1,t) + f_{\text{sup}}(z_1,t)
\Big),
\notag
\end{align}
where the force densities are given by
\begin{align*}
f_{\text{Br}}(z_1,t) &= -\frac{\partial \rho(z_1,t)}{\partial z_1} ,
\\
f_{\text{ext}}(z_1,t) &= - 
\rho(z_1,t)\frac{\partial V_{\text{ext}}(z_1,t)}{\partial z_1} ,
\\
f_{\text{ad}}(z_1,t) &= 2\pi\! 
\int_{z_1-1}^{z_1+1}\! dz_2 \, (z_1-z_2) \, \rho^{(2)}_{\text{ad}}(z_1,z_2,r_2^{*},t) ,
\\
f_{\text{sup}}(z_1,t) &= 2\pi\! 
\int_{z_1-1}^{z_1+1}\! dz_2 \, (z_1-z_2) \, \rho^{(2)}_{\, \text{sup}}(z_1,z_2,r_2^{*},t),
\end{align*}
and where $r_2^{*}=\!\sqrt{1-(z_1-z_2)^2}$ is the value of the cylindrical radius on the surface of the excluded volume 
sphere ($r_{12}\!=\!1$). 
A detailed derivation of the one-dimensional integrals required for 
calculation of $f_{\text{ad}}$ and $f_{\text{sup}}$, which are specific to the hard-sphere system, can 
be found in Appendix C of Reference \cite{tschopp3}. 
If the external field represents a 
hard-wall, then it is convenient to delete 
$f_{\text{ext}}$ from equation \eqref{one body planar} and instead represent the wall using a zero-flux 
boundary condition, as outlined in the text following equation \eqref{one-body exact}.

Treatment of the two-body equation of motion \eqref{two body adiabatic} requires more care than the one-body equation discussed above. 
We recall that to deal with the two-body correlation functions in planar geometry we have chosen $z_1$
to lie along the axis of our cylindrical coordinate 
system. 
This choice is essential for efficient solution of the 
inhomogeneous OZ equation \eqref{oz} (see Appendix 
\ref{appendix adiabatic two-body functionals}),  
but has the side-effect of introducing an asymmetry 
into the notation (but not the physics) of the equation of motion for the two-body density. 
While the coordinates $z_1$, $z_2$ and $r_2$ can be freely 
varied, the coordinate $r_1$ is constrained to the value $r_1\!=\!0$. 
This seemingly frustrates a straightforward calculation of derivatives 
with respect to $r_1$ and thus the evaluation of the 
Laplacian, $\nabla^{\,2}_{\vec{r_1}}$, of the superadiabatic two-body density.
This difficulty can, however, be avoided by exploiting the 
translational and rotational symmetry of the two-body density in the plane perpendicular to the $z$-axis. We define the divergence of the two-body current at 
coordinate $\vec{r}_2$ according to
\begin{align} \label{div_twobody_current}
\zeta_2(z_1,z_2,r_2,t) &=
\frac{1}{r_2} \frac{\partial}{\partial r_2} 
r_2 \frac{\partial}{\partial r_2} \rho^{(2)}_{\text{\,sup}}(z_1,z_2,r_2,t) \\
&+ \frac{\partial^2}{\partial z_2^2} \, \rho^{(2)}_{\text{\,sup}}(z_1,z_2,r_2,t) 
\notag\\
&+ \frac{\partial}{\partial z_2} \Bigg(\, \rho^{(2)}(z_1,z_2,r_2,t) \, \frac{\partial V_{\text{ext}}(z_2,t)}{\partial z_2}\,\Bigg)  
\notag\\
&- \frac{\partial}{\partial z_2}\Bigg(\;
\rho^{(2)}_{\text{ad}}(z_1,z_2,r_2,t) \, \frac{\partial V_{\text{ad}}(z_2,t)}{\partial z_2} \;\Bigg)\,. 
\notag
\end{align}
The direct interaction contribution to 
$\zeta_2$  is singular for hard-spheres and has thus been omitted from the right-hand side of equation \eqref{div_twobody_current}. 
This term is instead handled by imposing a zero-flux boundary condition on the excluded volume 
contact shell.  
This procedure ensures that the core condition on the two-body density is satisfied at all times 
and is essential 
for a correct description of structural packing effects. 
After some consideration of the geometrical sketch 
shown in Fig.~\ref{sketch_planar}, it becomes apparent that 
the divergence of the two-body current at coordinate 
$\vec{r}_1$ can be obtained from the information 
already contained in the function $\zeta_2$, 
namely
\begin{align} \label{symmetry relation}
\zeta_1(z_1,z_2,r_2,t) 
=
\zeta_2(z_2,z_1,r_2,t) \,.
\end{align}
We can thus rewrite equation 
\eqref{two body adiabatic} as
\begin{align} \label{twobody_eom_planar}
&\frac{\partial \rho^{(2)}(z_1,z_2,r_2,t)}{\partial t} = 
-
\zeta_2(z_1,z_2,r_2,t) 
-
\zeta_2(z_2,z_1,r_2,t)\,, 
\end{align}
which is the form suitable for the numerical 
study of hard-sphere systems in planar geometry.

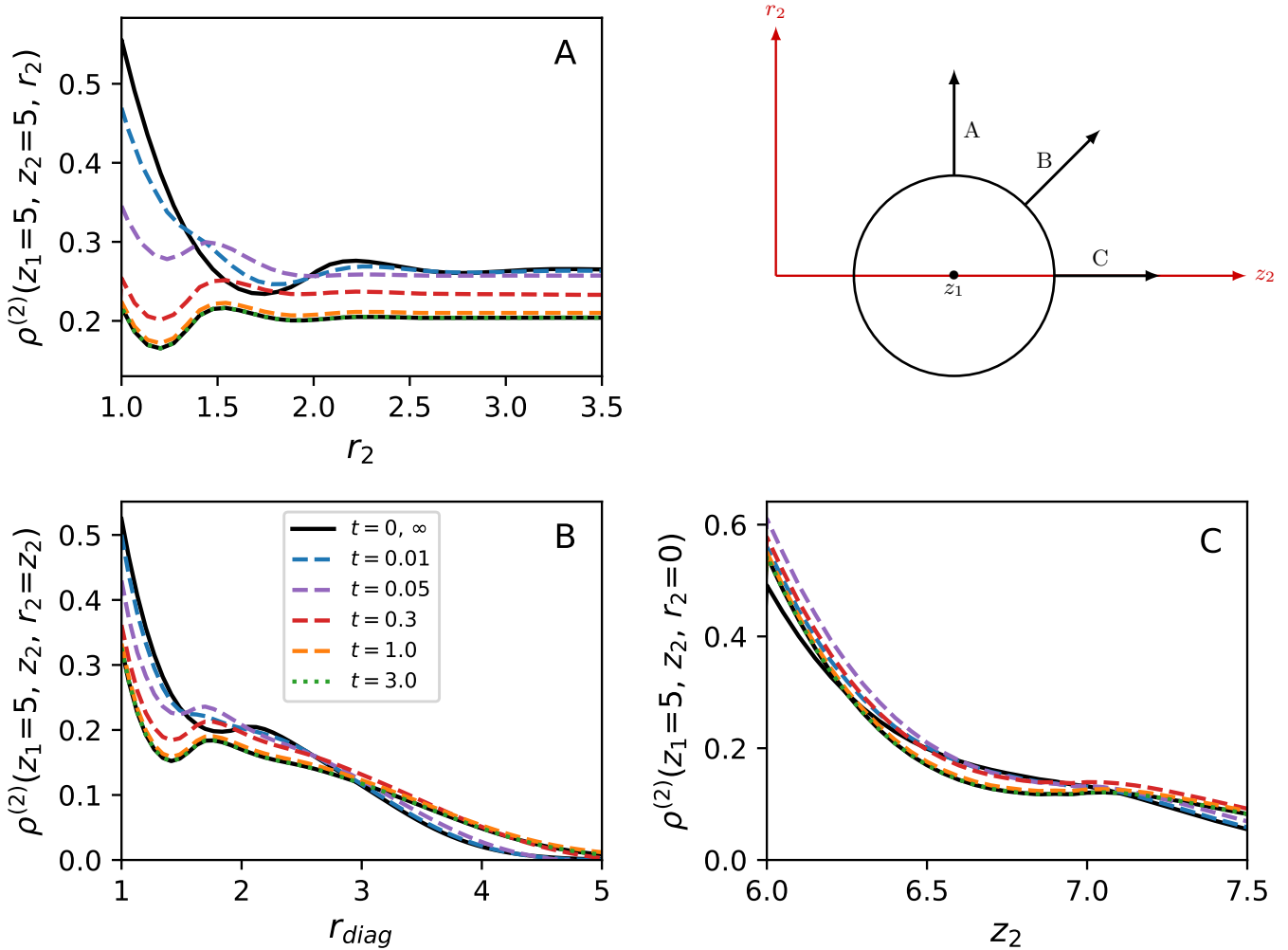
\begin{figure*}
\hspace*{0cm}
\begin{minipage}[t]{1\linewidth}
\begin{tikzpicture}
\node[inner sep=0pt] (fig10) at (0,0) {\includegraphics[width=1\linewidth]{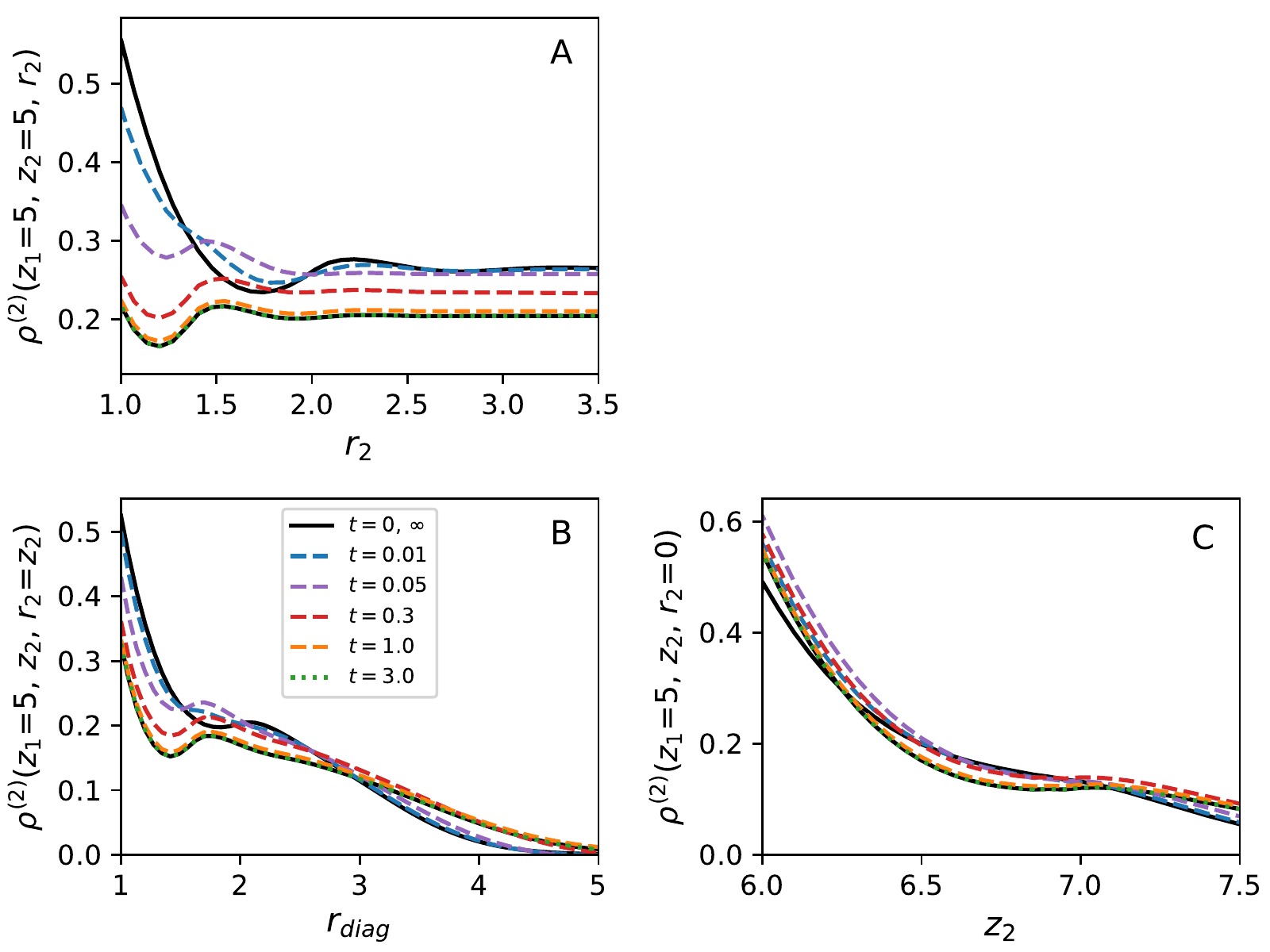}};
    %
\coordinate (origine) at (0+1.925,0+2.85);
\coordinate (z1) at (2.5+1.925,0+2.85);
\coordinate (z2 end) at (6.6+1.925,0+2.85);
\coordinate (r2 end) at (0+1.925,3.5+2.85);
\coordinate (vertical arrow start) at (2.5+1.925,1.4+2.85);
\coordinate (vertical arrow end) at (2.5+1.925,2.9+2.85);
\coordinate (diagonal arrow start) at (3.4899494936611664+1.925,0.9899494936611664+2.85);
\coordinate (diagonal arrow end) at (4.550609665440987+1.925,2.0506096654409873+2.85);
\coordinate (horizontal arrow start) at (3.9+1.925,0+2.85);
\coordinate (horizontal arrow end) at (5.4+1.925,0+2.85);
\coordinate (vertical label) at (2.75+1.925,2.05+2.85);
\coordinate (diagonal label) at (3.77027958+1.925,1.62027958+2.85);
\coordinate (horizontal label) at (4.55+1.925,0.25+2.85);
\node[below] (z1 label) at (z1) {$z_1$};
\node[right, bostonuniversityred] (z2 label) at (z2 end) {$z_2$};
\node[above, bostonuniversityred] (r2 label) at (r2 end) {$r_2$};
\draw[->, >=latex, line width=1.1, black] (vertical arrow start) -- (vertical arrow end);
\draw[->, >=latex, line width=0.9, bostonuniversityred] (origine) -- (z2 end);
\draw[->, >=latex, line width=0.9, bostonuniversityred] (origine) -- (r2 end);
\draw[->, >=latex, line width=1.1, black] (diagonal arrow start) -- (diagonal arrow end);
\draw[->, >=latex, line width=1.1, black] (horizontal arrow start) -- (horizontal arrow end);
\draw[line width=1] (z1) circle (40pt);
\draw[black] (z1) node {$\bullet$};
\draw[black] (vertical label) node {A};
\draw[black] (diagonal label) node {B};
\draw[black] (horizontal label) node {C};
\end{tikzpicture}
\end{minipage}
\caption{
{\bf Test case.}  
Time-evolution of the two-body density (calculated from 
equation \eqref{two body adiabatic}) for fixed  external potential and one-body density.  
The external potential is the harmonic trap \eqref{harmonic_potential} with $A\!=\!0.5$ and 
$z_1\!=\!z_0\!=\!5$, and the one-body density is held fixed to the 
corresponding equilibrium profile with average particle number $\langle N\rangle\!=\!2$. 
We take the equilibrium two-body density of a system with 
trap amplitude $A\!=\!0.75$ as the initial condition. 
As this initial condition is inconsistent with the fixed density and external potential we observe the relaxation of 
the two-body density towards the equilibrium state corresponding 
to a trap amplitude $A\!=\!0.5$. 
The geometrical sketch indicates the direction along which 
the two-body density is shown in the various panels. 
\label{trap_fig1}
}
\end{figure*}

\subsection{Numerical implementation}
\label{sec:numerics}

Numerical solution of the coupled 
equations \eqref{one body planar}, \eqref{div_twobody_current} and \eqref{twobody_eom_planar} is performed on a 
discrete grid. 
The gridpoints in the $z$-direction are 
evenly spaced with grid spacing $dz$, whereas the 
gridpoints in the $r_2$-direction are determined by 
the zeros of the Bessel function $J_0$. This 
choice facilitates numerical solution of the 
inhomogeneous OZ equation \eqref{oz} required 
for evaluation of all adiabatic quantities 
(see Appendix \ref{Appendix:solving OZ}). 
For details of the discretization we refer the reader to References \cite{tschopp2} and 
\cite{Lado}. 
In the following we outline the main points of our numerical algorithm for situations in which 
the system is initially in an equilibrium state before being subjected to a time-dependent external potential.

\begin{itemize}
\item Choose the starting external potential, $V^{\text{start}}_{\text{ext}}$, 
and the average number of particles per unit area, $\langle N \rangle\!=\!
\int_{-\infty}^{\infty}dz \rho(z)$.

\item Using the well-known Rosenfeld functional to generate the two-body direct correlation function (Appendix \ref{Appendix:FMT}) we solve the equilibrium force-DFT, defined 
by equations \eqref{YBG 1},\eqref{oz},\eqref{rho2 definition} and \eqref{c2 functional} to obtain  
the initial one- and two-body densities. 
In this equilibrium state  we have that 
$V_{\text{ad}}\!=\!V^{\text{start}}_{\text{ext}}$
and $\rho^{(2)}\!=\!\rho^{(2)}_{\text{ad}}$.
%
\item The external potential becomes time-dependent for $t\!>\!0$. 
This initiates the dynamics as the system tries to relax to the equilibrium state corresponding to the instantaneous external field. The external and adiabatic potentials now differ, $V_{\text{ad}}\!\neq\!V_{\text{ext}}$, and the right-hand side of \eqref{div_twobody_current} becomes nonzero.
\item Evaluate $\zeta_2$ from equation \eqref{div_twobody_current} for all values of $z_1$, which can be treated as a parameter. 
It is here that the zero flux condition is imposed on the excluded volume contact shell. 
This can be efficiently implemented using 
the method of `ghost points', as detailed in Appendix \ref{Appendix:ghost points}. 
\item

Use equation \eqref{twobody_eom_planar} and 
standard Euler integration to time-step the two-body density and then employ the definition \eqref{rho sup} 
to construct $\rho^{(2)}_{\,\text{sup}}$. 
\item Substitute $\rho^{(2)}_{\text{ad}}$ and 
$\rho^{(2)}_{\,\text{sup}}$ into the 
exact one-body equation of motion 
\eqref{one body planar} and use standard Euler integration to time-step the one-body density. 
\item Use the new one-body density to construct the adiabatic state using equations 
\eqref{oz},\eqref{rho2 definition} and \eqref{c2 functional}. Substitute the obtained $\rho^{(2)}_{\text{ad}}$ into equation 
\eqref{V ad} to calculate the adiabatic one-body force. 
\item Return to the third step of this list and 
continue time-stepping the coupled dynamics.
\end{itemize}

\subsection{Test case: dynamics of the two-body density}
\label{sec:test}

Superadiabatic-DDFT consists of a pair of coupled differential equations for the one- and two-body densities, namely \eqref{one-body exact} and 
\eqref{two body adiabatic}. 
The one-body equation of motion \eqref{one-body exact} has the same structure as the force-DDFT \eqref{force-DDFT}. 
In contrast, the structure of the equation of motion for the two-body density \eqref{two body adiabatic} is completely new.   
The double-dynamics generated by the full 
superadiabatic-DDFT scheme is highly nontrivial, 
due to the strong coupling between the one- and two-body densities.  
To get a feeling for the dynamical behaviour of the two-body density in isolation we consider first an artificial test case for which the 
one-body density, as well as all related adiabatic quantities, are held fixed. 

For this case, shown in Fig. \ref{trap_fig1}, we consider three-dimensional hard-spheres confined by a one-dimensional harmonic 
potential, 
\begin{equation}\label{harmonic_potential}
V_{\text{ext}}(z) 
= A\,(z-z_0)^{2}, 
\end{equation}
with amplitude $A$ and a minimum 
located at $z\!=\!z_0$. The one-body density and its related adiabatic quantities are calculated for a harmonic trap with $A\!=\!0.5$, $z_0\!=\!5$ and average number of particles per unit area $\langle N\rangle\!=\!2$.
The starting (full) two-body density, $\rho^{(2)}$, is set equal to the adiabatic two-body density obtained for a harmonic trap of amplitude
$A\!=\!0.75$.
As this initial condition is inconsistent with the fixed density and external potential it will force the two-body density to relax towards the equilibrium state corresponding 
to the trap of amplitude $A\!=\!0.5$. 
In the top right panel of Fig.\ref{trap_fig1} a 
geometrical sketch indicates the direction along which the two-body density results are shown. 
In all panels the coordinate $z_1\!=\!5$. The chosen times are given in the legend of panel B.
The full black lines in all panels indicate the equilibrium two-body density for both of the chosen trap amplitudes. 

Panel A follows the variation of $\rho^{(2)}$ 
in the cylindrical radial direction, $r_2$, for which the external field remains constant. 
We choose $z_2\!=\!z_1\!=\!5$. 
As the time increases we observe the relaxation of $\rho^{(2)}$ from the initial, nonequilibrium state, to the final equilibrium state of the system with $A\!=\!0.5$. By time $t\!=\!3$ the 
two-body density has fully relaxed. (This observation applies also in panels B and C.) 
Because the external field is constant the two-body density attains a constant value as 
$r_2$ increases. 
Panel B shows the relaxation of the two-body density
along the diagonal where the coordinates $|z_2-z_0|\!=\!r_2$, i.e. as a function of the distance $r_{\text{diag}}\!=\!\sqrt{(z_2-z_0)^2 + r^2_2}$.
We observe that $\rho^{(2)}$ relaxes most rapidly close to the excluded volume contact-shell (at $r_{\text{diag}}\!=\!1$). 
For large values of $r_{\text{diag}}$ the two-body density decays to zero as the external harmonic potential \eqref{harmonic_potential} 
grows in magnitude. 
Finally, panel C shows the relaxation of $\rho^{(2)}$ as a function of $z_2$ for the fixed value $r_2\!=\!0$. For the same reason as in panel B the two-body density decays to zero. However, this occurs more rapidly here because the external field is steeper in the $z_2$-direction than in the $r_{\text{diag}}$-direction.

This test case demonstrates that equation 
\eqref{two body adiabatic} generates the relaxation behaviour expected of the two-body density and that our numerical algorithms capture this correctly. 
In the following we will release the constraint 
on the one-body density and address the full 
superadiabatic dynamics of the system.
The time-evolution of the one- and two-body densities is then given by the coupled equations 
\eqref{one body planar} and \eqref{twobody_eom_planar}.

\subsection{Results: superadiabatic dynamics}
\label{sec:results}

\begin{figure*}
\hspace*{-0.6cm}
\includegraphics[width=0.65\linewidth]{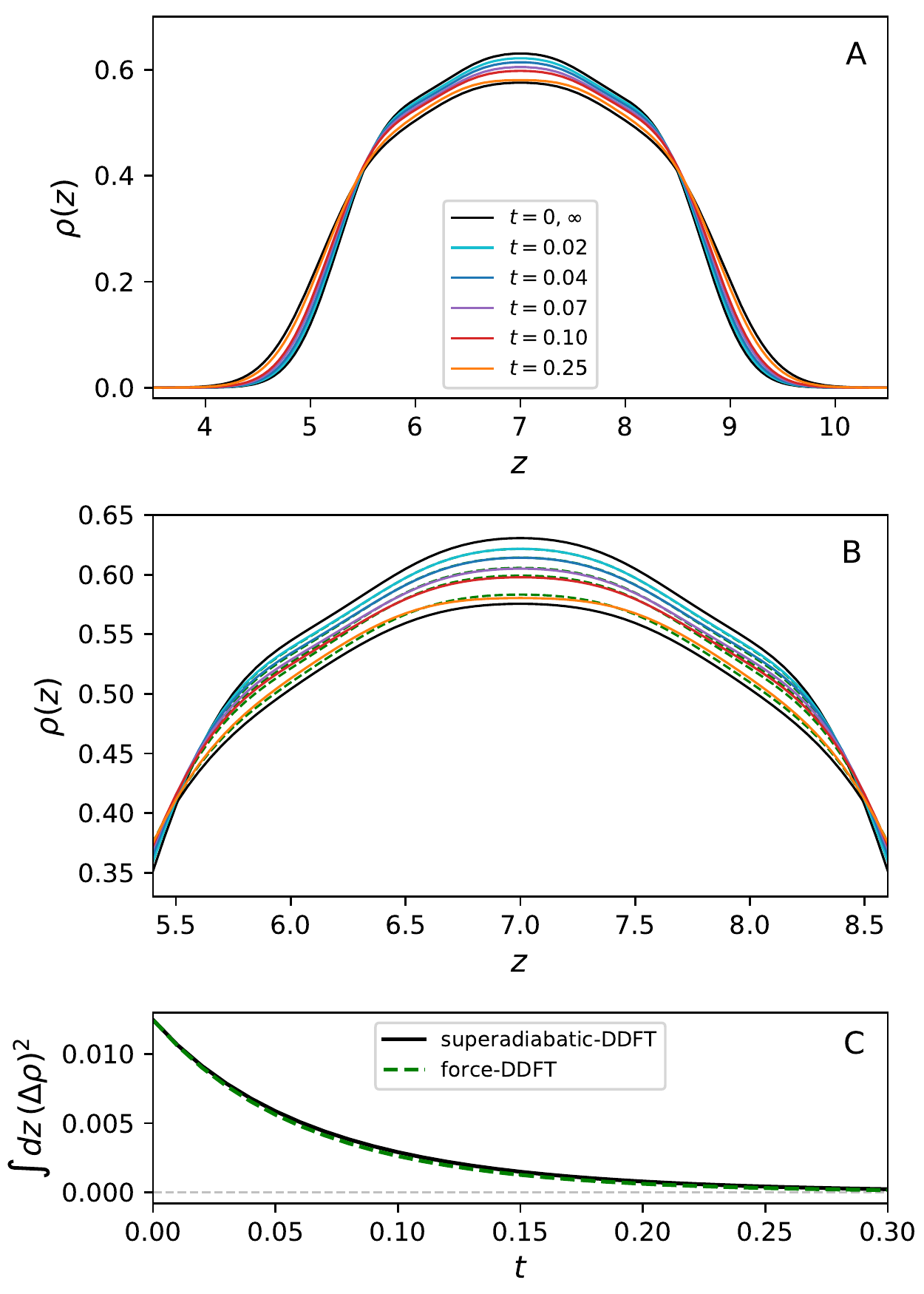}
\caption{
{\bf Harmonic trap.} Time-evolution of the density following a discontinuous 
change in the trap amplitude from $A\!=\!1.5$ to $1.1$ at time $t\!=\!0$.
Panels A and B show the density profiles for different time-steps. Panel A shows the full profiles, whereas panel B shows only a zoom at the peak position. The black lines are the density profiles at equilibrium, obtained with force-DFT. The solid color lines show the dynamical results obtained with superadiabatic-DDFT. In panel B the dashed green lines show the density profiles obtained with force-DDFT for comparison.
Superadiabatic- and force-DDFT start with the same equlilibrium curve and then deviate from each other as time passes. Note that the final equilibrium curve is the same for both schemes and thus both will relax to the same profile at long times.
Panel C shows the squared difference between the density profiles and the target equilibrium profile at each time-step, integrated over the $z$-coordinate. 
This panel shows that the superadiabatic-DDFT (solid black line) relaxes slower than the force-DDFT (dashed green line).
\label{trap_fig3}
}
\end{figure*}

\begin{figure*}
\hspace*{-0.6cm}
\includegraphics[width=0.63\linewidth]{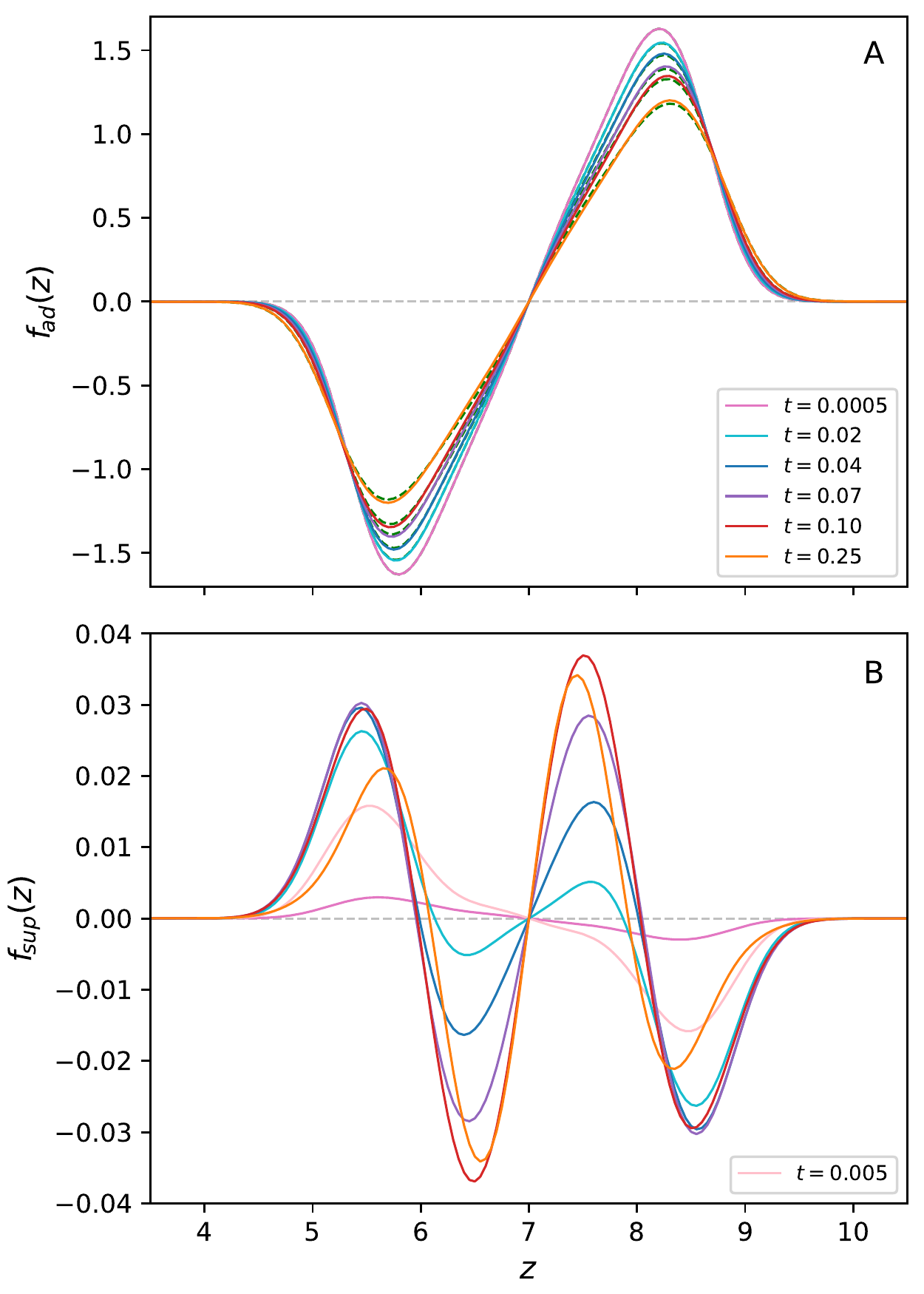}
\caption{
{\bf Harmonic trap.} Companion plot to 
Fig. \ref{trap_fig3} showing the time-evolution of the adiabatic and superadiabatic force densities
during the relaxation of the system. 
Panel A compares the adiabatic force from superadiabatic-DDFT (solid colored lines) with that from force-DDFT (dashed green lines). 
The slight differences arise because of the feedback between $f_{\text{sup}}$ and $f_{\text{ad}}$ during the evolution of the one-body density. 
Panel B Shows the superadiabatic force density at the same time-steps as in panel A with the addition of a curve for $t\!=\!0.005$. The curve initially starts at a constant value of zero, then increases in amplitude until reaching a peak around $t\!=\!0.07$ from which it decreases until completely vanishing for long times. This behaviour is to be expected since superadiabatic forces are absent in equilibrium states and can only develop during dynamical processes. 
\label{trap_fig4}
}
\end{figure*}

{\bf Harmonic trap.} 
We now consider once again the external potential given by equation \eqref{harmonic_potential}, but this time with an initial amplitude $A\!=\!1.5$ which is then switched to amplitude $A\!=\!1.1$ for times $t\!\ge\!0$. The relaxation of the three-dimensional hard-sphere system is shown in Figs. \ref{trap_fig3} and \ref{trap_fig4}. 
We have chosen to first investigate the density relaxation in switching from one \textit{harmonic trap} to another, anticipating that the force-DDFT will already make accurate predictions 
and that the superadiabatic-DDFT will only provide small corrections. 
The aim here is to investigate how even small superadiabatic forces affect the structure of the one-body density, through their presence in equation \eqref{one body planar}. (We will later 
consider a case where the superadiabatic forces have a more dramatic influence.)

In Fig. \ref{trap_fig3} we show the time-evolution of the density following this discontinuous 
change in the trap amplitude.
Panels A and B show the density profiles at different times, specified in the legend of panel A. The full profiles are shown in panel A, whereas panel B focuses on the peak position. The black lines in both panels show the density profiles at equilibrium, obtained with force-DFT. The solid color lines show the dynamical results obtained with superadiabatic-DDFT. In panel B the additional dashed green lines show the density profiles obtained with force-DDFT for comparison.
At equilibrium, both superadiabatic- and force-DDFT have identical density profiles. Starting from the same curve, the density profiles deviate more and more from each other as time increases. We observe that the superadiabatic-DDFT density profiles exhibit more complex structure than those from force-DDFT, since the two-body density in superadiabatic-DDFT respects the core-condition at all times and must thus evolve through a more realistic sequence of packing configurations (force-DDFT is not subject to this constraint). This new mechanism is also responsible for the slower relaxation time of superadiabatic-DDFT. 
In the long-time limit, both superadiabatic- and force-DDFT relax to the same density profile, as the superadiabatic forces vanish (see the comments on the following Fig. \ref{trap_fig4} for more details).
Finally, to illustrate the aforementioned relaxation-time, panel C shows the squared difference between the nonequilibrium density profiles and the target equilibrium profile at each time-step, integrated over the $z$-coordinate, i.e.
\begin{equation*}
\int_{-\infty}^{\infty} dz\, 
\left(\Delta\rho(z,t)\right)^2,
\end{equation*}
where $\Delta\rho(z,t)\!=\!\rho(z,t)\!-\!\rho(z,t\!\rightarrow\!\infty)$.
The superadiabatic values are shown in black, whereas the force values are shown in dashed green.
This panel shows explicitly that the superadiabatic-DDFT indeed relaxes slower than the force-DDFT and, as anticipated, this effect is rather small for the present case.

The companion plot, Fig. \ref{trap_fig4}, shows the time-evolution of the adiabatic and superadiabatic force densities
during the relaxation of the system. 
Panel A compares the adiabatic force from superadiabatic-DDFT (solid colored lines) with that from force-DDFT (dashed green lines). 
The slight differences arise because of the feedback between $f_{\text{sup}}$ and $f_{\text{ad}}$ during the evolution of the one-body density. 
Note that the adiabatic force density remains finite in the final equilibrium state. 
Panel B shows the superadiabatic force density at the same times as in panel A, with the addition of a curve for $t\!=\!0.005$. The superadiabatic force density initially starts at zero, then increases in amplitude until reaching a maximum at around $t\!=\!0.07$ from which point it then decreases gradually to zero at long times.
The growth and subsequent decay of the superadiabatic force density makes clear that this is a truly nonequilibrium quantity.
Note that the maximal amplitude of the superadiabatic force density is much smaller than the amplitude of the adiabatic force density. This small contribution is responsible for the (expected) small differences between superadiabatic- and force-DDFT in the present case. In the following we will investiguate a different external potential for which the superadiabatic force density is of much larger relative amplitude. This leads to more dramatic structural changes in the one-body density and a stronger slow-down in the relaxation to equilibrium.\\

\begin{figure*}
\hspace*{-0.6cm}
\includegraphics[width=0.7\linewidth]{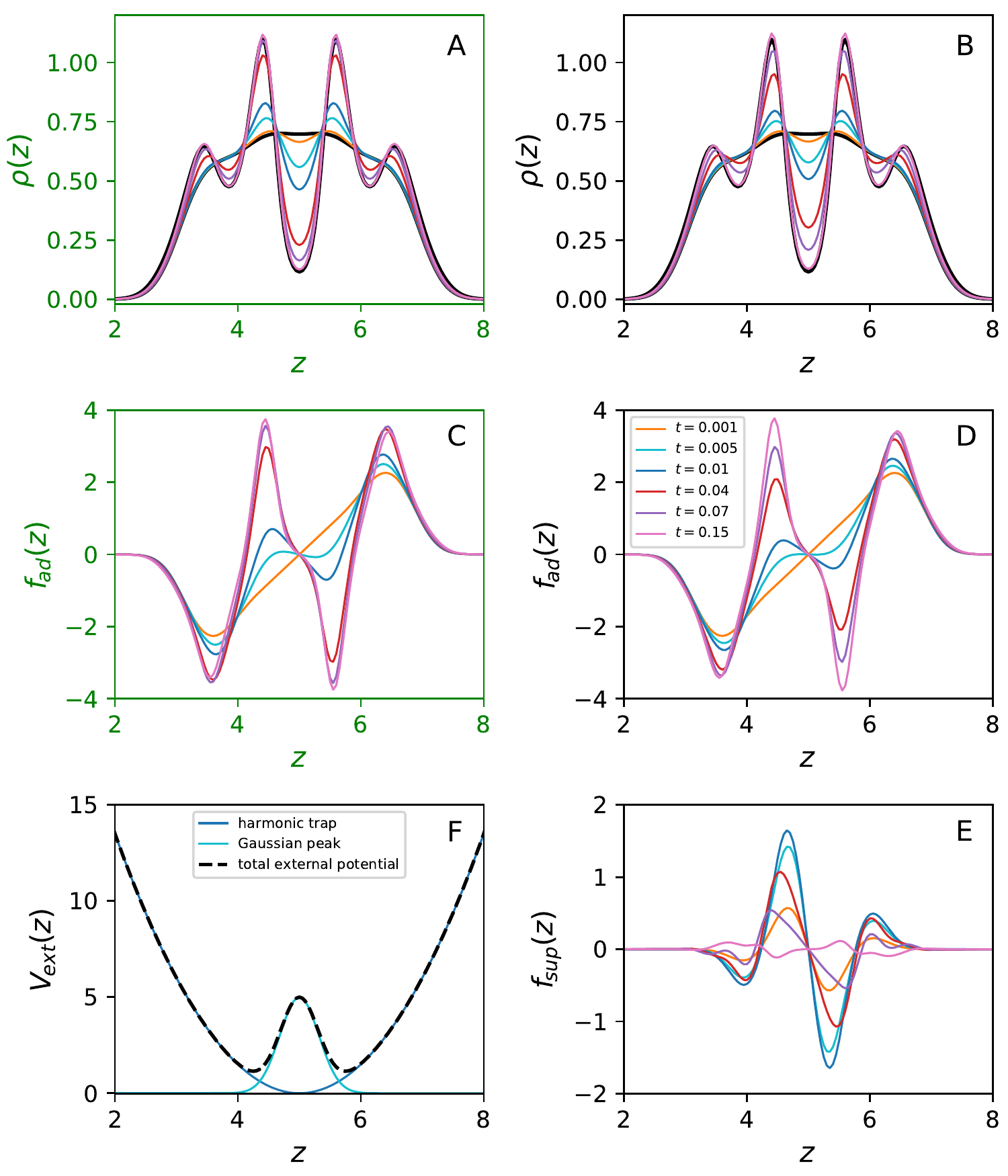}
\caption{
{\bf Harmonic trap plus Gaussian peak.} In panels A and B we show one-body density profiles obtained from force- and superadiabatic-DDFT, respectively, following a discontinuous change in the external potential (as illustrated in panel F) at time $t\!=\!0$. 
The initial external potential is a harmonic trap of amplitude $A\!=\!1.5$, shown in dark blue in panel F. At time $t\!=\!0$ we switch on an additional Gaussian peak (as described in the main text), which is shown in light blue in the same panel. The resulting external potential for $t\!>\!0$ is then given by the dashed black curve. 
Focusing again on panels A and B, we observe a clear difference in the time-evolution predicted by the force- and the superadiabatic-DDFT. 
See the companion Figure \ref{foxy_fig6} for 
a more detailed comparison. 
Panels C, D and E focus on the adiabatic and superadiabatic force densities. 
In contrast to the simple harmonic trap investigated in Figures \ref{trap_fig3} and 
\ref{trap_fig4}, the adiabatic force densities 
shown in panels C (force-DDFT) and D (superadiabatic-DDFT) are quite distinct. The differences are due to the large amplitude of the superadiabatic force density, shown in panel E.   Its maximum amplitude is of the order of half  the amplitude of the adiabatic force density, shown in panel D. As already explained in the case of the simple harmonic trap, the superadiabatic force density grows from zero at time $t\!=\!0$, increases until reaching a peak in amplitude and then decays to zero at long times.
Note that the superadiabatic force density exhibits complex structure and a rather unpredictible time evolution. This
is to be contrasted with the more straightforward evolution of the adiabatic force densities.
\label{foxy_fig5}
}
\end{figure*}

\begin{figure*}
\hspace*{-0.6cm}
\includegraphics[width=0.95\linewidth]{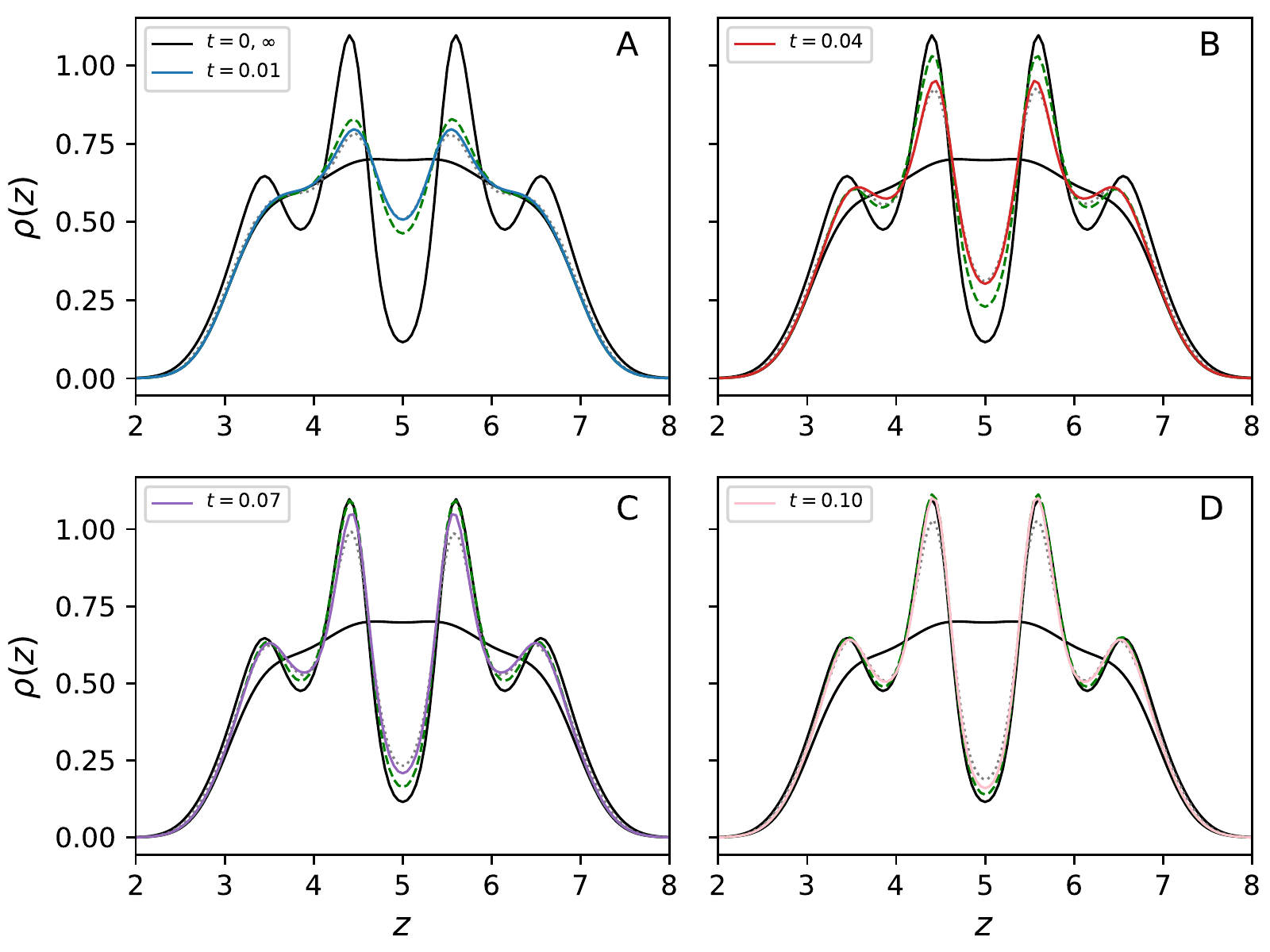}
\caption{
{\bf Harmonic trap plus Gaussian peak.} Companion plot to 
Fig. \ref{foxy_fig5} showing results for the one-body density at four different times, one per panel. In each panel the solid colored lines are the density profiles obtained with superadiabatic-DDFT. They can be compared with those from force-DDFT, shown as dashed green lines. Brownian dynamics simulation data are shown as gray dotted lines. The equilibrium density profiles are given by the solid black lines as a guide for the eye.
In panel A the superadiabatic density profile is in excellent agreement with the simulation data, whereas the force-DDFT curve already presents some clear deviation.
In panel B the superadiabatic profile remains in very good agreement with the simulations, while  the deviation of force-DDFT from simulation increases even more than in the preceding panel.
In panel C both superadiabatic- and force-DDFT profiles deviate from the simulation data. Nevertheless the superadiabatic curve is in much better agreement with simulation than the force-DDFT.
Finally, in panel D we approach the long-time limit where both superadiabatic- and force-DDFT are expected to converge to the same equilibrium profile (which is not in perfect agreement with simulation, as a result of the approximate free energy functional used).
For all times considered the superadiabatic-DDFT provides a globally improved account of the simulation data and relaxes much more slowly than force-DDFT.
\label{foxy_fig6}
}
\end{figure*}

{\bf Harmonic trap plus Gaussian peak.} 
We now consider the following external potential:
\begin{equation}\label{foxy_potential}
V_{\text{ext}}(z,t) 
= A\,(z-z_0)^{2} 
+ 
B(t)\,e^{-\alpha(z-z_0)^2},
\end{equation}
where $A\!=\!1.5$, $\alpha\!=\!5$, $z_0\!=\!5$ and the time-dependent Gaussian coefficient is given by
\begin{equation*}
  B(t) =
    \begin{cases}
      0 & \text{for\;} t<0,\\
      5 & \text{for\;} t\ge0. 
    \end{cases}       
\end{equation*}
In this case we set the average number of particles per unit area $\langle N\rangle\!=\!2.5$ such that the initial state for $t\!<\!0$ is rather densely packed. 
Switching on the repulsive Gaussian peak at $t\!=\!0$ pushes the particles out from the center and towards the edges of the trap, thus forcing them to undergo strongly correlated collective motion in order to arrive at the final equilibrium profile. This displays a pronounced 
second peak reflecting the particle layers which develop on each side of the trap. 
In early studies of one-dimensional hard-rods \cite{Marconi98} it was found that situations for which the system has to evolve through highly correlated states are poorly described by standard DDFT. 
If the value of $\langle N\rangle$ is sufficiently large, then 
applying the external potential \eqref{foxy_potential} presents a somewhat analogous situation for three-dimensional hard-spheres (with the benefit of being free of the 
ensemble differences which complicate one-dimensional studies). 
We thus expect the superadiabatic forces to be of a magnitude comparable to that of the adiabatic forces and that they will make a significant contribution to the relaxation of the one-body density.
Figs. \ref{foxy_fig5} and \ref{foxy_fig6} display results for the situation described above (where the external field is illustrated in Fig. \ref{foxy_fig5} panel F), for a system of three-dimensional hard-spheres.

In  panels A and B of Fig. \ref{foxy_fig5} we show one-body density profiles obtained from force- and superadiabatic-DDFT, respectively.
We observe a clear difference in the time-evolution predicted by the force- and the superadiabatic-DDFT which confirms that the superadiabatic forces are playing a significant role the dynamics.   
Panels C, D and E focus on the adiabatic and superadiabatic force densities at various times during the relaxation.  
In contrast to the simple harmonic trap investigated in Figs. \ref{trap_fig3} and \ref{trap_fig4}, the adiabatic force densities 
shown in panels C (force-DDFT) and D (superadiabatic-DDFT) are quite distinct. 
Despite the fact that the two theories predict very similar curves 
for $f_{\text{ad}}$ at short times, differences in both form and amplitude clearly emerge already by time $t\!=\!0.01$. 
It can be seen that the $f_{\text{ad}}$ from superadiabatic-DDFT 
(panel D) lags behind that predicted by force-DDFT (panel C). 
This is due to the feedback of the superadiabatic force density, 
shown in panel E. 
Starting from zero in equilibrium, $f_{\text{sup}}$ grows in amplitude during the time evolution until a time around $t\!=\!0.01$ 
and then decreases to zero again as the new equilibrium state is approached. 
The form of $f_{\text{sup}}$ becomes increasingly complex as time increases and does not have any simple relation to the adiabatic force density curve. 
Moreover, we observe that the maximum amplitude of $f_{\text{sup}}$ is of the order of half the amplitude of $f_{\text{ad}}$ and thus exerts a strong influence on the time evolution of the one-body density.   

In Fig. \ref{foxy_fig6} we show results for the one-body density at four different times, one per panel. In each panel the solid colored lines are the density profiles obtained with superadiabatic-DDFT. These can be compared with those from force-DDFT, shown as dashed green lines. Brownian dynamics simulation data are shown as gray dotted lines. The equilibrium density profiles are given by the solid black lines as a guide for the eye.
In panel A the superadiabatic density profile is in excellent agreement with the simulation data, whereas the force-DDFT curve already deviates significantly.
In panel B the superadiabatic profile remains in very good agreement with the simulation data, while the error in force-DDFT continues to grow. 
In panel C both superadiabatic- and force-DDFT profiles deviate from the simulation data. Nevertheless the superadiabatic curve is in much better agreement with simulation than the force-DDFT.
Finally, in panel D the long-time limit is approached. Both superadiabatic- and force-DDFT converge to the same equilibrium profile, namely the profile predicted by force-DFT. 
Note that the equilibrium force-DFT profile is not in perfect agreement with simulation as a result of the approximate free energy functional used (as discussed in reference \cite{tschopp3}).
For all times considered the superadiabatic-DDFT provides a globally improved account of the simulation data. 

In this example, not only does the one-body density from superadiabatic-DDFT relax much more slowly than in force-DDFT but, even more importantly,
the shape of the curves are in far better agreement with the Brownian dynamics simulation data. The details of the relaxation are captured more accurately because the superadiabatic-DDFT correctly implements 
the core-condition on the two-body density. 
This yields a physically reasonable description of the structural relaxation in the system, which then feeds back into the time-evolution 
of the one-body density.

\section{Discussion} \label{discussion}

In this paper we have developed and implemented a first-principles superadiabatic-DDFT for inhomogeneous fluids out-of-equilibrium.
By explicitly treating the dynamics of the two-body correlation functions the theory generates superadiabatic forces directly from the microscopic interparticle interactions. 
The importance of these nonequilibrium forces 
for a quantitatively accurate description of the one-body density is demonstrated by our numerical implementation of the theory for hard-spheres in a time-dependent planar potential. 
By treating correctly the constraints on 
the two-body density the superadiabatic-DDFT incorporates the physics of structural 
relaxation, essential for a realistic treatment 
of dynamics in dense, strongly correlated systems.  
Our new theory thus solves the long-standing problem of how to go beyond standard adiabatic DDFT and opens up new avenues for future research. 

The standard DDFT \eqref{ddft standard form} and force-DDFT \eqref{force-DDFT} represent alternative methods to implement adiabatic dynamics on the one-body level, which become 
equivalent when the excess Helmholtz free energy functional is known exactly \cite{tschopp3}. 
When using an approximate free energy functional the force-DDFT does not yield any quantitative improvement over standard DDFT, but does provide 
a more natural starting point for the development of a superadiabatic theory. 
This is achieved, in the new superadiabatic-DDFT, by replacing the adiabatic two-body density with 
the dynamic two-body density determined by solving equation \eqref{two body adiabatic}. 
The fact that the dynamic two-body density takes 
time to `catch up' with the time-evolving one-body density, rather than being instantaneously equilibrated, is the mechanism by which superadiabatic forces are generated within our theory, see equation \eqref{f_sup}. 
We argue that employing the one-body direct correlation function, $c^{(1)}$, to approximate the average interaction force has for a long time blocked progress in developing theories beyond leading-order adiabatic DDFT. 
Working with the two-body density enables us to develop a more explicit and intuitive connection to the forces 
acting in the system (via equations \eqref{f_ad} and \eqref{f_sup}) and makes the progression to a higher-order theory appear rather natural. Although implementation of the superadiabatic-DDFT does come with increased computational demands, these are in fact quite modest by todays standards and should not be regarded as a barrier 
to application of the theory. 

Our approach to going beyond standard DDFT involves truncating the exact hierarchy of dynamical equations for the 
$n$-body density at second-order and only then 
applying the adiabatic approximation. 
This naturally raises the more general, and fundamental, question of whether adiabatic 
truncation at higher-orders would lead to a 
convergent sequence of approximations.
In this sense, the present work can also be regarded as a numerical evidence which supports the convergence of the adiabatically truncated series. 
We mention that an analogous question was considered long-ago for the case of equilibrium 
bulk fluids. 
The Born-Green equation is a well-known equation 
for the bulk two-body density, which is derived from the exact YBG2 equation \eqref{YBG 2} by applying 
the Kirkwood superposition approximation to 
the three-body density \cite{Hansen06, mcquarrie}. 
Ree {\it et al.} \cite{Ree1,Ree2} 
investigated whether improved results could be obtained by considering the next equation in the heirarchy and applying a 
superposition approximation to the four-body density. 
Solving the resulting coupled pair of equations 
for the bulk two- and three-body densities indeed resulted in much improved performance. 
Although this equilibrium scheme clearly differs from the present dynamical theory, it nevertheless provides a further example that heirarchy truncation \textit{can be} a powerful tool for the development of improved approximations in liquid-state theory. 

The numerical results we have presented for two 
different time-dependent potentials show that the superadiabatic-DDFT relaxes slower than the force-DDFT. 
This behaviour can be understood by considering 
the time-evolution of the two-body density within each of these approaches. 
In the case of force-DDFT the two-body density is 
assumed to be instantaneously equilibrated at each time step to the one-body density.  
This generates an unphysical trajectory of the two-body density through function-space, which does not respect the structural constraints induced by the interparticle interactions (the core condition, in the case of hard-spheres). 
In contrast, these constraints are explicitly 
treated in superadiabatic-DDFT, via the appearance of the gradient of the bare pair-potential in equation \eqref{two body adiabatic}.  The correlated rearrangement of particles in 
a strongly interacting system, often referred to as `structural relaxation', is a 
key concept in studies of slow dynamics in glasses and gels \cite{CatesSoftBook}. 
As mentioned previously, we could 
in principle eliminate the two-body density from the coupled equations of superadiabatic-DDFT and arrive at a single equation of motion for the one-body density. 
This equation would then involve a time 
integral over the history of the system. 
The complexity of the superadiabatic scheme does not allow this to be carried out in practice, but memory effects are incorporated implicitly when solving the pair of coupled differential equations \eqref{one-body exact} and \eqref{two body adiabatic}.  
The superadiabatic-DDFT method of using 
time-local one- and two-body densities to treat
structural relaxation can be 
constrasted with the mode-coupling theory of 
{\it bulk} fluids \cite{modecoupling}.  
In the latter approach structural relaxation and slow dynamics in the spatially homogeneous bulk are captured by an explicit  memory kernel in the dynamical equation for a two-time 
density auto-correlation function.  

The application of superadiabatic-DDFT presented in this work demonstrates that the inclusion of 
superadiabatic forces leads to a quantitative improvement over standard DDFT for relaxational dynamics (see Fig. \ref{foxy_fig6}). 
For systems subject to external driving 
we anticipate that the superadiabatic-DDFT will 
 capture nonequilibrium phenomena inaccessible to standard DDFT. 
In the presence of a time-dependent external velocity field, $\vec{v}(\vec{r},t)$, the many-body Smoluchowski equation \eqref{smoluchowski} is generalized to the following 
\begin{multline}\label{Smolochowski flow}
\frac{\partial P(\rv^N\!,t)}{\partial t}
=
\sum_{i=1}^{N}
D_0 \nabla_{\vec{r}_i} \cdot \Big( P(\rv^N\!,t)\big(
\nabla_{\vec{r}_i} \ln (P(\rv^N\!,t)) \Big. \\ \Big. 
+ \nabla_{\vec{r}_i} \beta U(\rv^N\!,t)
\big) \Big) 
- \nabla_{\vec{r}_i} \cdot \Big(\vec{v}(\vec{r}_i,t)P(\rv^N\!,t)\Big).
\end{multline}
Integration of this equation over $N\!-\!1$ particle coordinates yields equation \eqref{one-body exact} with the additional term
\begin{equation}\label{extra term 1}
- \nabla_{\vec{r}_1}\! \cdot \Big(\vec{v}(\vec{r}_1,t)\rho(\vec{r}_1,t)\Big),
\end{equation}
while integration over $N\!-\!2$ particle coordinates yields equation \eqref{two-body exact} with the additional term
\begin{equation}\label{extra term 2}
- \sum_{i=1,2} \nabla_{\vec{r}_i}\! \cdot \Big(\vec{v}(\vec{r}_i,t)\rho^{(2)}(\vec{r}_1,\vec{r}_2,t)\Big).
\end{equation}
Note that when integrating equation \eqref{Smolochowski flow} one has to treat correctly the boundary terms. If these do not reduce to zero, then the equations above cannot be used as stated.
Consider the case of an infinite bulk system subject to a steady shear flow in the $x$-direction and shear gradient in the $z$-direction, such that 
$\vec{v}(\vec{r})\!=\!\dot{\gamma}\,z\,\unit_x$, where $\dot{\gamma}$ is the shear rate.
In this case the one-body density is a constant and is not affected by the shear flow. 
Although the one-body equation of motion becomes irrelevant, superadiabatic-DDFT gives access to the equation of motion for $\rho^{(2)}$ which is able to capture the flow distorted two-body density. 
From this, one can calculate the interaction part of the virial stress tensor and 
thus access rheological quantities such as the 
shear viscosity (see, for example, references \cite{BraderReview,BraderKruger,BraderScacchi}). 
If we now add a hard-wall boundary in the $xy$-plane to the sheared system described above, 
then the coordinates of the particles are restricted to the half-space with $z\!>\!0$. 
In this case the one-body density is no longer a constant, but becomes a function of both position and shear-rate (because the particles in the vicinity of the hard-wall are forced to move past each other in the direction of flow).  
Standard DDFT does not capture the shear-rate dependence, 
since the additional flow term \eqref{extra term 1} vanishes \cite{BraderKruger}.
However, in contrast, the term given by \eqref{extra term 2} does remain and generates 
a nonzero superadiabatic force in the $z$-direction due to the shear-induced distortion of the two-body density.
This then assures the desired $\dot{\gamma}$-dependency of the one-body density profile.
Incorporating external flow in the superadiabatic-DDFT provides the gateway to the theoretical study of inhomogeneous fluid rheology from first principles.
One could then envisage applications to relevant phenomena such as shear-induced migration and shear-banding where the external flow field has a dramatic influence on the one-body density profile.
However, the implementation of superadiabatic-DDFT for systems under flow presents a technical challenge, because the cylindrical symmetry of the two-body correlation functions considered in the present work no longer holds. Investigations in this direction will be the subject of future work.

The difficulty in treating systems under external flow stems from the reduction of symmetry and the consequent increase in complexity of the two-body correlation functions. 
Leaving aside these geometrical complications, let us now consider generalisation to more complicated systems.
Both DFT and DDFT can be readily generalized to treat mixtures of different particle species. 
This is not only useful for extending the range of systems which can be investigated, but 
can also be exploited to implement particle `tagging' strategies to access the internal 
dynamics of fluids in equilibrium. 
A key quantity of interest in equilibrium dynamics is the van Hove function \cite{Hansen06,mcquarrie} 
describing the motion of a tagged particle away from its initial position as well as the 
corresponding dynamic rearrangement of the surrounding particles. 
By treating the density of the tagged particle (species 1) and the density of the others 
(species 2) separately, DDFT enables calculation of the self and distict parts of the van Hove 
function. 
Implementations of this strategy for hard-spheres have revealed clearly the limitiations of 
standard DDFT in describing slow dynamics and particle `caging' effects at high 
density \cite{dynamic_test,roth_vanhove,pft_test}. 
Thus, it would be of considerable interest to 
revisit this type of problem using superadiabatic-DDFT.



\acknowledgments
We thank H. Vuijk for providing the Brownian dynamics simulation data shown in Fig. \ref{foxy_fig6}.

\appendix

\section{Standard DFT} \label{appendix potential-DFT}

The central object within standard DFT is the grand potential functional \cite{Evans79,Evans92}
\begin{align*}
\Omega[\rho] = F^{\,\text{id}}[\rho] + F^{\,\text{exc}}[\rho] 
- \int \!d\rv \big( \mu - V_{\text{ext}}(\rv) \big)\rho(\rv), 
\end{align*}
where $\mu$ is the chemical potential and $V_{\text{ext}}(\rv)$ is the external potential. 
The Helmholtz free energy of the ideal gas is given exactly by	
\begin{align*}
F^{\,\text{id}}[\rho]=k_BT\int\!d\rv\, \rho(\rv)\big(\ln(\rho(\rv))-1\big),
\end{align*}
where the (physically irrelevant) thermal wavelength has been set equal to unity. 
The excess Helmholtz free energy functional, $F^{\,\text{exc}}
$, encodes the interparticle 
interactions and usually has to be approximated.
The grand potential satisfies the variational condition
\begin{align}
\label{variation omega}
{\frac{\delta  \Omega[\rho]}{\delta \rho(\rv)}}\bigg\rvert_{\rho_{\text{eq}}}=0,
\end{align}
which then generates the Euler-Lagrange equation 
for the equilibrium density
\begin{align} \label{EL}
\rho_{\text{eq}}(\rv)=\rm{e}^{ -\beta\left(V_{\text{ext}}(\rv) - \mu - k_BT \, c_{\text{eq}}^{(1)}(\rv; [\rho_{\text{eq}}])\right)}.
\end{align}
The one-body direct correlation function is defined according to 
\begin{align} \label{c1 definition}
c_{\text{eq}}^{(1)}(\rv; [\rho_{\text{eq}}]) = - {\frac{\delta \beta F^{\,\text{exc}}[\rho]}{\delta\rho(\rv)}}\bigg\rvert_{\rho_{\text{eq}}}
\end{align}
and acts as an effective external field, generated by interparticle interactions.

\section{Standard DDFT} \label{appendix standard DDFT}

The starting point when developing a closed dynamical theory 
for the one-body 
density is the exact equation \eqref{one-body exact}. 
We begin by making the adiabatic approximation to the force 
integral
\begin{align} \label{appendix B force integral}
\int d \vec{r}_2 \, &\rho^{(2)}(\vec{r}_1,\vec{r}_2,t) \,\nabla_{\vec{r}_1} \beta \phi(r_{12}) 
\notag \\
&\; \approx \int d \vec{r}_2 \,  \rho^{(2)}_{\text{ad}}(\vec{r}_1,\vec{r}_2,t)  \nabla_{\vec{r}_1} \beta \phi(r_{12}), 
\end{align}
where the adiabatic two-body density is defined in equation 
\eqref{rho2 ad}. 
Using this approximation in the exact expression \eqref{one-body exact} generates the 
force-DDFT \eqref{force-DDFT}. 
This can be expressed in a more compact form by applying the equilibrium YBG equation \eqref{YBG 1} to reexpress the three-body integral 
\begin{align} \label{appendix B ybg rewrite}
&\int d \vec{r}_2 \,  \rho^{(2)}_{\text{ad}}(\vec{r}_1,\vec{r}_2,t)  \nabla_{\vec{r}_1} \beta \phi(r_{12})
\\
& 
\qquad
\stackrel{\text{YBG}}{=}\, -\nabla_{\vec{r}_1} \rho(\vec{r}_1,t) - \rho(\vec{r}_1,t) \nabla_{\vec{r}_1} \beta V_{\text{ad}}(\vec{r}_1) ,
\notag
\end{align}
where the gradiant of the adiabatic potential is defined by \eqref{V ad}. 
Substitution of \eqref{appendix B ybg rewrite} into 
\eqref{force-DDFT} yields the following alternative expression of the force-DDFT
\begin{align} \label{appendix B force-DDFT}
\!\!\frac{1}{D_0} \frac{\partial \rho(\vec{r},t)}{\partial t} =& \nabla_{\vec{r}} \!\cdot\! \Big( 
\rho(\vec{r},t) \nabla_{\vec{r}}\, \beta 
\Big( V_{\text{ext}}(\vec{r},t) - V_{\text{ad}}(\vec{r},t)\Big)
\Big). 
\end{align}
Equation \eqref{appendix B force-DDFT} shows 
that changes in the one-body density are driven by the 
difference between the real and adiabatic potentials. 
This difference vanishes in equilibrium and we recover the 
YBG equation \eqref{YBG 1}. 
Note that we have arrived at the adiabatic 
equation of motion \eqref{appendix B force-DDFT} without 
any Reference to the variational principle \eqref{variation omega} or 
the resulting Euler-Lagrange equation \eqref{EL}.

The transition from the force- to the standard DDFT comes 
if we calculate the adiabatic potential in \eqref{appendix B force-DDFT} using the Euler-Lagrange equation \eqref{EL}, 
rather than the force-based expression \eqref{V ad}. 
In this case we have
\begin{equation*}
\nabla_{\vec{r}} \,\beta V_{\text{ad}}(\vec{r},t) = 
-\nabla_{\vec{r}} \ln \rho(\vec{r},t) + \nabla_{\vec{r}}\, c_{\text{ad}}^{(1)}(\rv,t), 
\end{equation*}
where the one-body adiabatic direct correlation function is 
defined according to 
\begin{equation*}
c_{\text{ad}}^{(1)}(\rv,t) 
\equiv
c_{\text{eq}}^{(1)}(\rv; [\rho(\vec{r},t)]),
\end{equation*}
with the equilibrium direct correlation function 
given by equation \eqref{c1 definition}. 
If we use equation \eqref{c1 definition} to explicitly introduce the Helmholtz free energy, then we obtain 
the more familiar form 
\begin{equation}\label{ddft standard form}
\frac{1}{D_0}\frac{\partial \rho(\rv,t)}{\partial t}
=
\nabla_{\vec{r}}\cdot\left(\rho(\rv,t)\,\nabla_{\vec{r}}
{\frac{\delta \beta \Omega[\rho]}{\delta\rho(\rv)}}\bigg\rvert_{\rho(\vec{r},t)}
\,\right).
\end{equation}
In equilibrium, this equation reduces 
to the variational principle of standard DFT \eqref{variation omega}. 

\section{Adiabatic two-body functionals}
\label{appendix adiabatic two-body functionals}

In the following we describe how 
to obtain the adiabatic two-body density as a functional of the one-body density for hard-spheres in planar geometry.

\subsection{Hard-sphere FMT}\label{Appendix:FMT}

The first quantity to calculate when taking a DFT approach to the two-body correlation functions is the inhomogeneous 
two-body direct correlation function, defined by equation 
\eqref{c2 functional}. 
This requires as input the excess Helmholtz free energy 
functional corresponding to the interaction potential 
of interest. 
Within the framework of FMT the excess Helmholtz free energy functional of the hard-sphere system can be written 
in the following form
\begin{align}\label{ros_fe}
\beta F^{\,\text{exc}}[\,\rho\,] = \int d\rv_1 \; \Phi \left( \left\lbrace n_{\alpha}(\rv_1) \right\rbrace  \right),
\end{align}
where the Helmholtz free energy density is a function of 
a set of weighted densities generated by convolving 
the one-body density profile with known weight functions 
\begin{equation}
n_{\alpha}(\rv_1) = \int d\rv_2 \; \rho(\rv_2)\, \omega_{\alpha}(\rv_1-\rv_2). 
\notag
\end{equation}
The weight functions, $\omega_{\alpha}$, are characteristic of the geometry of the hard spherical particles. 
Of the six weight functions, four are scalars 
\begin{align}
\omega_3(\rv)&=\Theta(R-r), \hspace*{0.5cm}
\omega_2(\rv)=\delta(R-r), \notag\\
\omega_1(\rv)&=\frac{\delta(R-r)}{4\pi R}, \hspace*{0.51cm}
\omega_0(\rv)=\frac{\delta(R-r)}{4\pi R^2}, \notag
\end{align}
and two are vectors (indicated by bold indices)
\begin{align}
\omega_{\bold 2}(\rv)&=\unit_{\rv}\,\delta(R-r),\hspace*{0.5cm}
\omega_{\bold 1}(\rv)=\unit_{\rv}\frac{\delta(R-r)}{4\pi R},
\notag
\end{align}
where $\unit_{\rv}=\rv/r$ is a unit vector. 
Although there exist various expressions for the Helmholtz 
free energy density, we choose to employ the original 
Rosenfeld form, given by \cite{Rosenfeld89,RothReview}
\begin{equation*}
\Phi = - n_0 \ln(1-n_3) + \frac{n_1 n_2 - {\bf n}_1 \cdot {\bf n}_2}{1-n_3} + \frac{n_2^3 
- 3 n_2 {\bf n}_2 \cdot {\bf n}_2}{24 \pi (1-n_3)^2}.
\end{equation*}
Using equation \eqref{ros_fe} in the definition \eqref{c2 functional} generates the following expression for the two-body direct correlation function
\begin{align}
c^{(2)}(\rv_1, \rv_2)
&=-\!\sum_{\alpha\beta}\int \!d\rv_3\, 
\omega_{\alpha}(\rv_{31})\,
\Phi''_{\alpha \beta}(\rv_3)\,
\omega_{\beta}(\rv_{32}),
\label{c_ros_fe}
\end{align}
where $\Phi''_{\alpha \beta}\!=\!\partial^2 \Phi/\partial n_{\alpha} \partial n_{\beta}$, the summation runs over all scalar and vector indices, and 
$\rv_{ij}=\rv_i-\rv_j$.  
For detailed descriptions of how to implement equation \eqref{c_ros_fe} in both planar 
and spherical geometries we refer the reader to Reference \cite{tschopp2}.

\subsection{Solving the OZ equation in planar geometry}\label{Appendix:solving OZ}

In planar geometry the two-body correlation functions are translationally and rotationally 
invariant in the plane perpendicular to the 
$z$-axis. 
Hankel transformation can be employed to exploit 
this symmetry and reduce 
the three-dimensional integral in the 
inhomogeneous OZ equation \eqref{oz} to a 
one-dimensional integral in the $z$-direction. 
The resulting equation is given by \cite{tschopp1} 
\begin{align}\label{oz_planar}
\overline{h}_{\text{ad}}(z_1,z_2,k) &= \overline{c}_{\text{ad}}^{\,(2)}(z_1,z_2,k) 
\\
&+ \int_{-\infty}^{\infty} \!dz_3 \;  
\overline{h}_{\text{ad}}(z_1,z_3,k)\,\rho(z_3)\,\overline{c}_{\text{ad}}^{\,(2)}(z_3,z_2,k)\,, 
\notag
\end{align}
where an overbar indicates a Hankel transformed quantity and $k$ is the absolute value of the two-dimensional wavevector $\vec{k}$. 
The forwards and backwards Hankel transforms of the two-body direct correlation function are defined according to
\begin{align*}
\overline{c}_{\text{ad}}^{\,(2)}(z_1,z_2,k) 
=&
2\pi\!\int_0^{\infty} \!dr\,r\,J_0(kr)\,
c_{\text{ad}}^{(2)}(z_1,z_2,r_2)\,,
\label{Hankel_forwards}\\
c_{\text{ad}}^{(2)}(z_1,z_2,r_2) 
=&
\frac{1}{2\pi}\!
\int_0^{\infty} \!dk\,k\,J_0(kr)\,
\overline{c}_{\text{ad}}^{\,(2)}(z_1,z_2,k)\,,
\end{align*}
where $J_0$ is a Bessel function. 
Analogous relations apply for the total correlation function.
For a given one-body density the Hankel-transformed inhomogeneous OZ equation 
\eqref{oz_planar} contains two unknown quantities, $\overline{h}_{\text{ad}}$ 
and $\overline{c}_{\text{ad}}$, and thus 
requires additional information to obtain a closed theory of the 
two-body correlation functions. 
There are two methods by which this can be achieved: 
(i) Supplement equation \eqref{oz_planar} with a closure relation, such as the Percus-Yevick 
approximation,
(ii) Employ the DFT definition 
\eqref{c2 functional} to obtain the two-body direct correlation function as a functional of the one-body density and then substitute this 
into \eqref{oz_planar}. 
Examples of these two strategies can be found in 
References \cite{tschopp1} and \cite{tschopp2}, 
respectively. 
For the case of hard-spheres treated with FMT 
a useful analytic formula for $\overline{c}_{\text{ad}}$ can be found in Reference 
\cite{tschopp2}, which is obtained by Hankel 
transformation of equation \eqref{c_ros_fe}. 
In the present work we wish to minimize the 
computational load of time-stepping the 
equations of superadiabatic-DDFT and thus, as in 
Reference \cite{tschopp3}, we choose to employ the second of these routes. The inhomogeneous OZ equation \eqref{oz_planar} is iteratively solved in Hankel-space for $\overline{h}_{\text{ad}}$ (since both $\rho$ and $\overline{c}_{\text{ad}}$ are known input quantities). When this has been found, we back transform the total two-body correlation function to real space and finally use the definition
\begin{equation*} 
\rho^{(2)}_{\text{ad}}(z_1, z_2, r_2) = \big(1+h_{\text{ad}}(z_1, z_2, r_2)\big) \rho_{\text{ad}}(z_1) \rho_{\text{ad}}(z_2)
\end{equation*} 
to obtain the desired adiabatic two-body density.

\section{Zero-flux boundaries and the ghost-point method}\label{Appendix:ghost points}


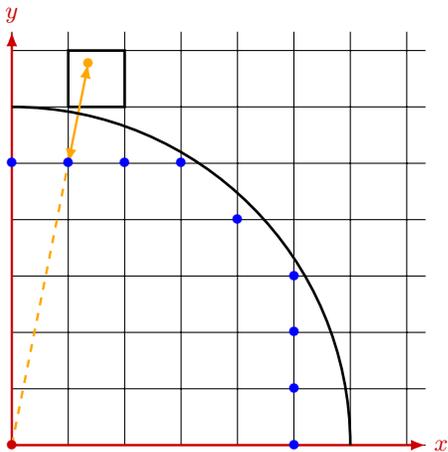
\begin{figure}
\begin{minipage}[t]{0.4\textwidth}
\hspace*{-0.5cm}
\begin{tikzpicture}
\coordinate (origine) at (0,0);
\coordinate (x end) at (5.5,0);
\coordinate (y end) at (0,5.5);
\coordinate (grid end) at (5.5,5.5);
\coordinate (ghost point 1) at (0,3.75);
\coordinate (ghost point 2) at (0.75,3.75);
\coordinate (ghost point 3) at (1.5,3.75);
\coordinate (ghost point 4) at (2.25,3.75);
\coordinate (ghost point 5) at (3,3);
\coordinate (ghost point 6) at (3.75,2.25);
\coordinate (ghost point 7) at (3.75,1.5);
\coordinate (ghost point 8) at (3.75,0.75);
\coordinate (ghost point 9) at (3.75,0);
\coordinate (lookup point 2) at (1.015,5.0745);;
\node[right, bostonuniversityred] (x label) at (x end) {$x$};
\node[above, bostonuniversityred] (y label) at (y end) {$y$};
\draw[step=0.75,black,thin] (origine) grid (grid end);
\draw[-, >=latex, line width=1, black] (0.75,4.5) -- (0.75,5.25) -- (1.5,5.25) -- (1.5,4.5) -- (0.75,4.5);
\draw[<->, >=latex, line width=0.9, chromeyellow] (ghost point 2) -- (lookup point 2);
\draw[->, >=latex, line width=0.9, bostonuniversityred] (origine) -- (x end);
\draw[->, >=latex, line width=0.9, bostonuniversityred] (origine) -- (y end);
\draw[line width=1] (4.5,0) arc (0:90:4.5);
\draw[dashed, >=latex, line width=0.9, chromeyellow] (origine) -- (ghost point 2);
\draw[blue] (ghost point 1) node {$\bullet$};
\draw[blue] (ghost point 2) node {$\bullet$};
\draw[blue] (ghost point 3) node {$\bullet$};
\draw[blue] (ghost point 4) node {$\bullet$};
\draw[blue] (ghost point 5) node {$\bullet$};
\draw[blue] (ghost point 6) node {$\bullet$};
\draw[blue] (ghost point 7) node {$\bullet$};
\draw[blue] (ghost point 8) node {$\bullet$};
\draw[blue] (ghost point 9) node {$\bullet$};
\draw[chromeyellow] (lookup point 2) node {$\bullet$};
\draw[bostonuniversityred] (origine) node {$\bullet$};
\end{tikzpicture}
\end{minipage}
\hspace*{2cm}
\caption{\textbf{The ghost-point construction on a two-dimensional square grid.} 
The zero-flux condition is applied on a circular boundary 
incommensurate with the two-dimensional grid. 
The blue circles indicate the ghost-points and the single 
orange circle gives an example of a lookup-point. 
The ghost-point and lookup-point are radially equidistant from the 
boundary. 
}
\label{sketch_ghost_points}
\end{figure}

Calculation of the two-body correlation functions in planar geometry is most conveniently performed in the cylindrical coordinate system. 
The two-body density then simplifies to a function of three scalar coordinates, $\rhotwo(z_1,z_2,r_2)$, in which we 
choose $z_1$ to be coincident with the cylindrical coordinate 
axis. 
In our numerical calculations of the two-body dynamics 
we regard the position $z_1$ as an external control parameter.
For a fixed value of $z_1$ the two-body density can be mapped onto a discrete two-dimensional grid with axes $z_2$ and $r_2$ (as shown in Fig. \ref{sketch_planar}). 
Calculating the time-evolution of $\rhotwo$ is thus reduced to the problem of solving a set of two-dimensional diffusion problems, one for each value of $z_1$. 

For hard-spheres the function 
$\rhotwo$ is identically zero within the excluded volume sphere (`core condition') and is thus subject to 
the zero-flux boundary condition $\hat{\vec{n}}\cdot \vec{j}\!=\!0$ on the sphere surface, where $\hat{\vec{n}}$ is the outwards-pointing normal unit-vector. 
The zero-flux condition would be easy to implement in spherical coordinates, however, numerical solution of the planar inhomogeneous OZ equation can only be efficiently performed using cylindrical coordinates. 
This leads to problems of grid incommensurability. 
To illustrate how to deal properly with this numerical difficulty we construct two simple examples, which capture the essential features.

Let us consider a two-dimensional ideal gas of infinite extent, which is excluded from a circular region of unit radius centred on 
the coordinate origin. 
For our chosen test-case the density of the gas is given by the time-dependent and circularly symmetric function $f(r_{\text{pol}},t)$, where the radial polar coordinate is related to the Cartesian coordinates according to $r_{\text{pol}}\!=\!\sqrt{x^2+y^2}$, and satisfies the zero-flux boundary condition on the unit circle. 
The diffusion equation in polar coordinates is given by
\begin{equation}
\frac{\partial}{\partial t} f(r_{\text{pol}},t) = -\frac{1}{r_{\text{pol}}} \frac{\partial}{\partial r_{\text{pol}}} \Bigg(r_{\text{pol}} \, j_{r_{\text{pol}}}(r_{\text{pol}}, t)\Bigg),
\label{diffusion_polar}
\end{equation}
where the radial current is
\begin{equation*}
j_{r_{\text{pol}}}(r_{\text{pol}}, t) = -D_0\,\frac{\partial f(r_{\text{pol}}, t)}{\partial r_{\text{pol}}},
\end{equation*}
and the boundary condition is imposed by setting $j_{r_{\text{pol}}}(r_{\text{pol}}\!=\!1, t)\!=\!0$. 
Numerical solution of equation \eqref{diffusion_polar} 
is straightforward for any initial choice of the function 
$f$ and this provides a benchmark for numerical solutions 
obtained using the (less convenient) Cartesian coordinate system. The diffusion equation in 
Cartesian coordinates is given by
\begin{multline}
\frac{\partial}{\partial t} f(x,y,t) = - \frac{\partial}{\partial x} j_{x}(x,y,t)
- \frac{\partial}{\partial y} j_{y}(x,y,t),
\label{cartesian_diffusion}
\end{multline}
where the currents in the $x$- and $y$-directions are
\begin{align}
j_{x}(x,y,t) &= -D_0\,\frac{\partial f(x,y,t)}{\partial x}, \label{currentx} \\
j_{y}(x,y,t) &= -D_0\,\frac{\partial f(x,y,t)}{\partial y}, \label{currenty}
\end{align} 
respectively.
As shown in Fig. \ref{sketch_ghost_points}, the points on a Cartesian grid do not in general lie on the circular boundary. To approximately impose the zero-flux condition we employ the method of ghost-points, as we describe below. 

Calculation of the partial derivatives given in \eqref{currentx} and \eqref{currenty} are performed using the central finite difference, for example in the $x$-direction
\begin{equation*}
\frac{\partial f(x,y,t)}{\partial x} \approx \frac{f(x\!+\!dx,y,t)-f(x\!-\!dx,y,t)}{2\, dx},
\end{equation*}
where $dx$ is the grid spacing.
For points closest to (but outside) the circular boundary, calculation of the central finite difference requires information about the value of $f$ at a neighbouring point inside the boundary.
The true function value at all inside points is zero, 
however, we can artificially attribute non-zero values to the set of first inner-points (referred to as `ghost-points') to guarantee that the central finite difference derivative in the radial direction yields zero exactly on the circular boundary, thus satisfying the zero-flux condition.  
To set the correct value at each ghost-point, it is necessary to find the corresponding `lookup-points' outside the circle via the geometrical construction shown in Fig. \ref{sketch_ghost_points}. The normal distance from a given ghost-point to the circular boundary is equal to $1-r_{\text{ghost}}$, where $r_{\text{ghost}}\!=\!\sqrt{x_{\text{ghost}}^2+y_{\text{ghost}}^2}$ is the length of the dashed-line. The distance from the origin to the lookup point location is therefore given by $r_{\text{lookup}} \!=\! 2-r_{\text{ghost}}$ and the corresponding $x$- and $y$-coordinates can then be calculated using simple trigonometry. 
The function value at the lookup point can be obtained with bilinear interpolation from the four surrounding 
grid-points, indicated by the bold-line square. 
This same value is then given to the corresponding ghost-point. 
When all ghost-point values are thus fixed the Cartesian derivatives \eqref{currentx} and \eqref{currenty} can be calculated at all outer-points in a way consistent with the 
desired boundary condition.   
From these current components the derivatives on the right-hand side 
of \eqref{cartesian_diffusion} can be calculated using central finite difference without further complication. 
The time-step is generated using standard Euler forward 
integration. 

We next consider the diffusion of an ideal gas in three dimensions, in which the gas is excluded from a spherical 
region centered on the coordinate origin. 
For our illustrative example we assume the density of the gas to be given by the spherically symmetric function $f(r_{\text{sph}},t)$. 
This situation is most naturally addressed in spherical coordinates, for which the diffusion equation takes the simple form
\begin{equation}
\frac{\partial}{\partial t} f(r_{\text{sph}}, t) = - \frac{1}{r_{\text{sph}}^2} \frac{\partial}{\partial r_{\text{sph}}} \Bigg( r_{\text{sph}}^2 \, j_{r_{\text{sph}}}(r_{\text{sph}}, t) \Bigg),
\label{diffusion_spherical}
\end{equation}
where the radial current is given by 
\begin{equation*}
j_{r_{\text{sph}}}(r_{\text{sph}}, t) = -D_0\,\frac{\partial f(r_{\text{sph}}, t)}{\partial r_{\text{sph}}},
\end{equation*}
and the boundary condition is imposed by setting $j_{r_{\text{sph}}}(r_{\text{sph}}\!=\!1, t)\!=\!0$. 
Equation \eqref{diffusion_spherical} is simple to solve 
for any initial choice of $f$ and thus provides useful 
reference data for numerical solutions obtained in 
other coordinate systems.

For the reasons already mentioned above we are compelled to formulate 
our diffusion problem in the cylindrical coordinate system. 
Due to the azimuthal symmetry the relevant coordinates are $z$ and the cylindrical radial coordinate $r_{\text{cyl}}$. 
The diffusion equation in cylindrical coordinates is given 
by
\begin{align}
\frac{\partial}{\partial t} f(z,r_{\text{cyl}},t) = 
&- \frac{1}{r_{\text{cyl}}} \frac{\partial}{\partial r_{\text{cyl}}} 
\Bigg(\! r_{\text{cyl}} \, j_{r_{\text{cyl}}}(z,r_{\text{cyl}},t) \! \Bigg) 
\label{diff_cyl}\\
&- \frac{\partial}{\partial z} j_{z}(z,r_{\text{cyl}},t) ,
\notag
\end{align}
where the current components in $z$- and 
$r_{\text{cyl}}\,$-directions are 
\begin{align*}
j_{z}(z,r_{\text{cyl}}, t) &= -D_0\,\frac{\partial f(z,r_{\text{cyl}}, t)}{\partial z} ,
\\
j_{r_{\text{cyl}}}(z,r_{\text{cyl}}, t) &= -D_0\,\frac{\partial f(z,r_{\text{cyl}}, t)}{\partial r_{\text{cyl}}} ,
\end{align*}
respectively.
When solving the OZ equation in cylindrical coordinates 
it is convenient (for numerical Hankel transformation) 
to work on a rectangular, but not evenly-spaced, grid.  
The $z$-direction is unproblematic, with regularly spaced 
grid-points. 
However, the points in the $r_{\text{cyl}}$ direction are determined by the zeros of the Bessel function, $J_0$. 
The resulting grid in the $z,r_\text{cyl}$ plane has 
points which do not lie on the excluded volume boundary. 
Implementation of the zero-flux boundary condition can 
be achieved using the ghost point method, in the same 
way as for our previous two-dimensional example.
As before, all numerical derivatives are performed using 
central finite difference.

Accurate numerical solution of the cylindrical diffusion equation \eqref{diff_cyl} is an essential pre-requisite 
for the solution of the second equation of 
superadiabatic-DDFT, equation  \eqref{two body adiabatic} in the main text.
Dealing with the fully interacting system naturally 
introduces additional numerical demands, namely the calculation of the adiabatic two-body density at each 
time-step. 
However, as far as the time-evolution is concerned, 
the only new consideration for solving equation 
\eqref{two body adiabatic} 
is that the center of the excluded volume sphere is centered at $z_1$, rather than at the coordinate origin. 
The methods presented in this appendix can thus be employed 
directly by using the coordinates $z_1$, $z_2$, $r_2$ in 
place of the $0$, $z$, $r_{\text{cyl}}$ used here. 
When developing our numerical algorithms we found it very useful to compare the time evolution of $f$ from numerical solution of the spherical equation \eqref{diffusion_spherical} with that of the cylindrical equation \eqref{diff_cyl}. 
An important requirement to be satisfied during the time-integration is that the normalization (spatial integral) of $f$ is conserved to an acceptable level of accuracy.

\bibliographystyle{apsrev4-2} 
\bibliography{paper4}




\end{document}